\documentclass[12pt]{article}%
\usepackage{amsmath}
\usepackage{amsfonts}
\usepackage{amssymb}
\usepackage{graphicx}%
\newtheorem{theorem}{Theorem}

\newtheorem{remark}[theorem]{Remark}

\begin{document}

\title{Eigenvalue problems, spectral parameter power series, and modern applications}
\author{Kira V. Khmelnytskaya$^{\text{1}}$, Vladislav V. Kravchenko$^{\text{2}}$,
\and Haret C. Rosu$^{\text{3}}$\\$^{\text{1}}${\small Faculty of Engineering, Autonomous University of
Queretaro, Queretaro, Mexico}\\$^{\text{2}}${\small Department of Mathematics, CINVESTAV del IPN, Campus
Queretaro, }\\{\small Apartado Postal 1-798, Arteaga \# 5, Col. Centro, Queretaro, Qro. }\\{\small 76001 Mexico, kravchenko@qro.cinvestav.mx}\\$^{\text{3}}${\small IPICyT, Instituto Potosino de Investigacion Cientifica y
Tecnologica,}\\{\small Apdo Postal 3-74 Tangamanga, 78231 San Luis Potos\'{\i}, Mexico,
hcr@ipicyt.edu.mx\bigskip}}
\date{}
\maketitle

\begin{abstract}
Our review is dedicated to a wide class of spectral and transmission problems
arising in different branches of applied physics. One of the main difficulties
in studying and solving eigenvalue problems for operators with variable
coefficients consists in obtaining a corresponding dispersion relation or
characteristic equation of the problem in a sufficiently explicit form.
Solutions of the dispersion relation are the eigenvalues of the problem. When
the dispersion relation is known the eigenvalues are found numerically even
for relatively simple problems with constant coefficients because even in
those cases as a rule the dispersion relation represents a transcendental
equation the exact solutions of which are unknown.

In the present review we deal with the recently introduced method of spectral
parameter power series (SPPS) and show how its application leads to an
explicit form of the characteristic equation for different eigenvalue problems
involving Sturm-Liouville equations with variable coefficients. We consider
Sturm-Liouville problems on finite intervals; problems with periodic
potentials involving the construction of Hill's discriminant and Floquet-Bloch
solutions; quantum-mechanical spectral and transmission problems as well as
the eigenvalue problems for the Zakharov-Shabat system. In all these cases we
obtain a characteristic equation of the problem which in fact reduces to
finding zeros of an analytic function given by its Taylor series. We
illustrate the application of the method with several numerical examples which
show that at present the SPPS method is the easiest in the implementation, the
most accurate and efficient. We emphasize that the SPPS method is not a purely
numerical technique. It gives an analytical representation both for the
solution and for the characteristic equation of the problem. 
This representation can be approximated by different numerical
techniques and for practical purposes constitutes a powerful numerical
method but most importantly it offers additional insight into the spectral and
transmission problems.

\end{abstract}


\section{Introduction}

Solution of second-order linear differential equations belongs to a classical
field of mathematics in which as an overwhelming majority of its users from
students to active researchers believe that everything that was possible to do
theoretically is essentially done, and whatever the analysts invent, in the
domain of practical solution of equations and related models anyway the
numerical discrete schemes more and more refined due to the massive efforts of
experts in numerical analysis and computer sciences will be more accurate and
efficient. For the good luck of mathematics and its applications a
considerable number of analysts as well as of advanced users of mathematical
methods in physics are aware of many strong limitations in applicability of
available numerical schemes. For examples, if the discrete spectrum of a
problem is not necessarily real, practically the whole machinery of advanced
numerical techniques does not apply bringing to the surface in fact one option
only: finite differences. The importance of this technique lies in its
universality. Nevertheless it usually gives way to many other approaches
whenever they become applicable. The main difficulty in finding complex
eigenvalues is due to the fact that the universally used shooting method works
if only there is a clear criterion for choosing every next shot. Meanwhile the
real numbers is an ordered set and zero is located always between a negative
and a positive outcomes of the corresponding shots, the complex plane does not
admit such a simple rule. There are many other situations (even when the
eigenvalues are real) when the shooting procedure finds considerable
difficulties and at the same time the method of finite differences is not
applicable at all. For example, when the spectral parameter participates in
the boundary conditions. Such situation in fact is more common in applications
than otherwise.

In the present review we discuss an approach developed in the last few years
and called the spectral parameter power series (SPPS) method. It is important
to notice that the SPPS method is not merely another numerical technique. On
the contrary, it is an analytical approach giving new analytical results and
at the same time lending itself to numerical calculation. The SPPS method
allows one to obtain two linearly independent solutions of the Sturm-Liouville
equation (in Section \ref{Sect Sol S-L} we specify the conditions imposed on
the coefficients)%
\begin{equation}
(pu^{\prime})^{\prime}+qu=\lambda ru \label{SLintro}%
\end{equation}
in the form%
\[
u_{1}(x)=\sum\limits_{k=0}^{\infty}a_{k}(x)\lambda^{k}\quad\text{and}\quad
u_{2}(x)=\sum\limits_{k=0}^{\infty}b_{k}(x)\lambda^{k}%
\]
where $\lambda$ is a spectral parameter and the series are uniformly
convergent. Such representations from time to time appear in mathematical
literature in different contexts. We mention here \cite[Sect. 10]{Bellman} and
\cite{DelsarteLions1956}. The main difference is the form in which the
coefficients $a_{k}$ and $b_{k}$, $k=0,1,\ldots$ are represented. In previous
works the calculation of the coefficients was proposed in terms of successive
integrals with the kernels in the form of iterated Green functions (see
\cite[Sect. 10]{Bellman}). This makes any computation based on such
representation difficult and less practical. Moreover, theoretical study of
the corresponding series and their properties becomes considerably more
complicated. We show that $a_{k}$ and $b_{k}$ can be calculated in terms of
the coefficients $p$ and $r$ of (\ref{SLintro}) as well as of a particular
solution of the equation $(pv^{\prime})^{\prime}+qv=0$. The obtained
representations of the coefficients $a_{k}$ and $b_{k}$ are relatively simple
and well suited both for theoretical estimates and for numerical computation
with a minimum of programmer's efforts required. Behind this representation of
the coefficients $a_{k}$ and $b_{k}$ there is one of the possible
factorizations of the operator $L=\frac{d}{dx}p\frac{d}{dx}+q$ sometimes
called the Polya factorization (see \cite{KelleyPeterson}).

The main advantage of the SPPS method is akin to that of such asymptotic
approaches as the WKB method - it allows one to work with an analytical
representation of the solution instead of a table of values delivered by a
numerical method. This is important from different points of view, often it
gives a new physical insight into the problem. The difference between the SPPS
and the asymptotic techniques lies in the fact that to apply the SPPS method
it is not necessary to assume the smallness or the largeness of the parameter
$\lambda$. For example, the WKB approximation can be efficiently applied to
(\ref{SLintro}) when $\lambda$ is sufficiently large which is quite useless
when an eigenvalue problem related to (\ref{SLintro}) is considered.

We show that 1) for solving initial value and boundary value problems the SPPS
method performs better or equal in comparison to purely numerical techniques;
2) it is highly advantageous when the solution is required for many different
values of the spectral parameter; 3) the SPPS method allows us to write down
an explicit form of the characteristic equations for many different spectral
problems which in practice reduces the spectral problem to finding zeros of a
corresponding analytic function given by its Taylor series. We emphasize that
the method is applicable in different situations when some other approaches
are unavailable (complex eigenvalues; $\lambda$-dependent boundary conditions,
etc.) The method is simple and can be introduced in mathematical courses for physicists.

In Section \ref{Sect Sol S-L}, we review the main results from \cite{KrCV2008}
and \cite{KrPorter2010} concerning the SPPS representation of the solutions of
(\ref{SLintro}) and show that even in solving initial and boundary value
problems this technique converted into a simple numerical algorithm is clearly
competitive when compared to standard routines for numerical integration of
linear ordinary differential equations. In Section \ref{SectSturm-Liouville}
we apply the SPPS method to Sturm-Liouville spectral problems with or without
the spectral parameter in the boundary conditions. Here together with some
results from \cite{KrPorter2010} we present new results concerning the
problems admitting complex eigenvalues. Section \ref{SectPeriodic} is
dedicated to the spectral problems for periodic potentials. We give an SPPS
representation for Hill's discriminant \cite{KiraRosu2010} and show how the
SPPS method allows one to construct the Bloch solutions of the problem. In
Section \ref{SectSpectral and Transmission} we consider two classical problems
of mathematical physics: the quantum-mechanical spectral problem and the
transmission problem. Following \cite{CKOR} we present a characteristic
(dispersion) equation equivalent to the eigenvalue problem for the
Schr\"{o}dinger operator with a potential which is an arbitrary continuous
function on a finite interval outside of which it is constant. We discuss some
numerical tests as well. The transmission problem is presented in the context
of electromagnetic wave propagation as the problem of calculation of the
reflection and transmission coefficients for a plane wave which is incident on
an inhomogeneous layer under an arbitrary angle of incidence. Following
\cite{CKKO2009} we discuss application of the SPPS method to this problem.
Section \ref{Sect Z-S} is dedicated to the eigenvalue problem for the
Zakharov-Shabat system. We present the dispersion equation \cite{KV2011} for
the problem for a real-valued, finitely supported potential and discuss its
practical application. Finally, in Section \ref{Sect Conclusions} we make some
concluding remarks.

This review is for the colleagues interested in all kinds of problems
involving solution of Sturm-Liouville type equations. We hope to attract more
attention to the SPPS approach which combines the possibility to work with
analytical representations of solutions and of characteristic equations of the
problems with simplicity, rapid convergence, and accuracy when used for
numerical computation.


\section{Spectral parameter power series representation for solutions of the
Sturm-Liouville equation\label{Sect Sol S-L}}

Let us consider the Sturm-Liouville equation
\begin{equation}
(pu^{\prime})^{\prime}+qu=\lambda ru~, \label{SL}%
\end{equation}
where $p$, $q$ and $r$ are complex valued functions and $\lambda$ is a complex
parameter. The following result \cite{KrPorter2010} gives us a convenient form
for a general solution of (\ref{SL}) as a spectral parameter power series.

\begin{theorem}
\label{ThSolSL} \cite{KrPorter2010} Assume that on a finite interval $[a,b]$,
equation
\begin{equation}
(pv^{\prime})^{\prime}+qv=0, \label{SL0}%
\end{equation}
possesses a particular solution $u_{0}$ such that the functions $u_{0}^{2}r$
and $1/(u_{0}^{2}p)$ are continuous on $[a,b]$. Then the general solution of
(\ref{SL}) on $(a,b)$ has the form
\begin{equation}
u=c_{1}u_{1}+c_{2}u_{2}~, \label{genmain}%
\end{equation}
where $c_{1}$ and $c_{2}$ are arbitrary complex constants,
\begin{equation}
u_{1}=u_{0}{\displaystyle\sum\limits_{k=0}^{\infty}} \lambda^{k}\widetilde
{X}^{(2k)}\quad\text{and}\quad u_{2}=u_{0}{\displaystyle\sum\limits_{k=0}%
^{\infty}} \lambda^{k}X^{(2k+1)} \label{gensol}%
\end{equation}
with $\widetilde{X}^{(n)}$ and $X^{(n)}$ being defined by the recursive
relations
\begin{equation}
\widetilde{X}^{(0)}\equiv1,\quad X^{(0)}\equiv1, \label{X0}%
\end{equation}%
\begin{equation}
\widetilde{X}^{(n)}(x)=\left\{
\begin{tabular}
[c]{ll}%
${\displaystyle\int\limits_{x_{0}}^{x}} \widetilde{X}^{(n-1)}(s)u_{0}%
^{2}(s)r(s)\,ds$, & $n$ \text{odd,}\\
${\displaystyle\int\limits_{x_{0}}^{x}} \widetilde{X}^{(n-1)}(s)\frac{1}%
{u_{0}^{2}(s)p(s)}\,ds$, & $n$ \text{even,}%
\end{tabular}
\ \ \ \ \ \right.  \label{Xtil}%
\end{equation}%
\begin{equation}
X^{(n)}(x)=\left\{
\begin{tabular}
[c]{ll}%
${\displaystyle\int\limits_{x_{0}}^{x}} X^{(n-1)}(s)\frac{1}{u_{0}^{2}%
(s)p(s)}\,ds$, & $n$ \text{odd,}\\
${\displaystyle\int\limits_{x_{0}}^{x}} X^{(n-1)}(s)u_{0}^{2}(s)r(s)\,ds$, &
$n$ \text{even},
\end{tabular}
\ \ \ \ \ \ \ \ \right.  \label{X}%
\end{equation}
where $x_{0}$ is an arbitrary point in $[a,b]$ such that $p$ is continuous at
$x_{0}$ and $p(x_{0})\neq0$. Further, both series in (\ref{gensol}) converge
uniformly on $[a,b]$.
\end{theorem}

For a detailed proof we refer to \cite{KrPorter2010}. It is based on some
simple observations. First of all the knowledge of a particular solution of
(\ref{SL0}) allows one to factorize the Sturm-Liouville operator $L=\frac
{d}{dx}p\frac{d}{dx}+q$ in the form $L=\frac{1}{u_{0}}\frac{d}{dx}%
\,p\,u_{0}^{2}\frac{d}{dx}\frac{1}{u_{0}}$, also known as the Polya
factorization \cite{KelleyPeterson} as mentioned before. This form is well
suited for establishing how the operator $\frac{1}{r}L$ acts on each member of
the series (\ref{gensol}). For example, $\frac{1}{r}L\left(  u_{0}%
\widetilde{X}^{(2k)}\right)  =u_{0}\widetilde{X}^{(2k-2)}$, $k\in\mathbb{N}$.
Analogously, $\frac{1}{r}L\left(  u_{0}X^{(2k+1)}\right)  =u_{0}X^{(2k-1)}$.

Considering the system of functions $\left\{  \varphi_{n}\right\}
_{n=0}^{\infty}$ defined as follows $\varphi_{0}=u_{0}$,%

\begin{equation}
\varphi_{n}(x)=\left\{
\begin{tabular}
[c]{ll}%
$u_{0}(x)X^{(n)}(x)$, & $n$ \text{odd,}\\
$u_{0}(x)\widetilde{X}^{(n)}(x)$, & $n$ \text{even,}%
\end{tabular}
\ \ \ \ \ \ \ \ \ \ \ \ \right.  \ \label{phik}%
\end{equation}
we find that $\frac{1}{r}L\varphi_{0,1}=0$ and $\frac{1}{r}L\varphi
_{n}=\varphi_{n-2}$, $n=2,3,\ldots$. These properties are characteristic for
the so-called $L$-bases (introduced in \cite{Fage}, unfortunately this
important book has not been translated into English), and hence formulas
(\ref{X0})-(\ref{X}) represent a practical way to calculate an $L$-basis. In
\cite{KrCMA2011} it was established that the system of functions $\left\{
\varphi_{n}\right\}  _{n=0}^{\infty}$ is complete in $L_{2}(a,b)$. For further
related properties we refer to \cite{KMoT}.

To establish the uniform convergence of the series in (\ref{gensol}) as well
as to get a rough but useful estimate for the velocity of their convergence it
is sufficient to observe that
\begin{equation}
\left\vert \widetilde{X}^{(2k)}\right\vert \leq\left(  \max\left\vert
ru_{0}^{2}\right\vert \right)  ^{k}\left(  \max\left\vert \frac{1}{pu_{0}^{2}%
}\right\vert \right)  ^{k}\frac{\left\vert b-a\right\vert ^{2k}}{\left(
2k\right)  !} \label{estimate}%
\end{equation}
(a similar inequality is available for $\left\vert X^{(2k+1)}\right\vert $ as
well). Thus, the series $%
{\displaystyle\sum\limits_{k=0}^{\infty}}
\lambda^{k}\widetilde{X}^{(2k)}$ is majorized by a convergent numerical series
$%
{\displaystyle\sum\limits_{k=0}^{\infty}}
\frac{c^{k}}{\left(  2k\right)  !}$ with $c=\left\vert \lambda\right\vert
\left(  \max\left\vert ru_{0}^{2}\right\vert \right)  \left(  \max\left\vert
\frac{1}{pu_{0}^{2}}\right\vert \right)  \left\vert b-a\right\vert ^{2}$.

It is worth noticing that from (\ref{gensol}) it is easy to obtain that
$u_{1}$ and $u_{2}$ satisfy the following initial conditions%
\begin{equation}
u_{1}(x_{0})=u_{0}(x_{0}),\qquad u_{1}^{\prime}(x_{0})=u_{0}^{\prime}(x_{0}),
\label{at01}%
\end{equation}%
\begin{equation}
u_{2}(x_{0})=0,\qquad u_{2}^{\prime}(x_{0})=\frac{1}{u_{0}(x_{0})p(x_{0})}.
\label{at02}%
\end{equation}

\begin{remark}
The possibility mentioned in the Introduction to represent solutions of
the Sturm-Liouville equation in the form of spectral parameter power series is
by no means a novelty, though it is not a widely used tool. In fact, besides
the work reviewed below we are able to mention only \cite[Sect. 10]{Bellman},
\cite{DelsarteLions1956} and the recent paper \cite{KostenkoTeschl}) and to
the best of our knowledge it was applied for the first time for solving spectral
problems in \cite{KrPorter2010}. The reason of this underuse of the SPPS lies
in the form in which the expansion coefficients were sought. Indeed, in
previous works the calculation of coefficients was proposed in terms of
successive integrals with the kernels in the form of iterated Green functions
(see \cite[Sect. 10]{Bellman}). However, this makes any computation based on such
representation difficult, less practical, and even proofs of the most basic
results like, e.g., the uniform convergence of the spectral parameter power
series for any value of $\lambda\in\mathbb{C}$ (established in Theorem
\ref{ThSolSL}) are not an easy task. For example, in \cite[p. 16]{Bellman} the
parameter $\lambda$ is assumed to be small and no proof of convergence is
given. Moreover, in \cite{CamposKr} a discrete analogue of Theorem
\ref{ThSolSL} together with some further applications to Jacobi operators were
established and it was pointed out that as well as in the continuous case the
SPPS representation for solutions of the Jacobi operators was considered as a
perturbation technique, however, even the situation with the convergence of
such series was not satisfactorily understood. We recommend the book \cite{Agarwal1992} where
the possibility of divergence of such series as those considered in the present
work is assumed. Due to the representation of the expansion coefficients
similar to (\ref{Xtil}), (\ref{X}), it was shown in \cite{CamposKr} that the
series are not only convergent but in the discrete case they are
actually finite sums.

The way of how the expansion coefficients in (\ref{gensol}) are calculated
according to (\ref{Xtil}) and (\ref{X}) is relatively simple and
straightforward. This is why the estimation of the rate of convergence of the
series (\ref{gensol}) presents no difficulty, see (\ref{estimate}). Another
crucial feature of the introduced representation of the expansion coefficients
in (\ref{gensol}) consists in the fact that as we repeatedly observe in
subsequent pages not only the expansion coefficients themselves also denoted
by $\varphi_{n}$ in (\ref{phik}) are necessary in solving different spectral
problems related to the Sturm-Liouville equation but also the formal powers
apparently obtained as a by-product of the recursive integration procedure
(\ref{Xtil}) and (\ref{X}), namely the functions $\widetilde{X}^{(2k+1)}$ and
$X^{(2k)}$, $k=0,1,2,\ldots$, which do not participate explicitly in the
representation of solutions (\ref{gensol}), naturally appear in dispersion
equations corresponding to the spectral problems. See (\ref{dispSL1}) and
(\ref{dispSL2}) for the dispersion equations equivalent to the classical
Sturm-Liouville eigenvalue problem, (\ref{D}) for the SPPS representation of
the Hill discriminant, (\ref{dispersionGeneral1}) for the dispersion equation
equivalent to the quantum-mechanical eigenvalue problem on the whole axis or
(\ref{cte}) for the dispersion equation equivalent to the Zakharov-Shabat
eigenvalue problem.

The consideration of formal powers (\ref{Xtil}) and (\ref{X}) as infinite
families of functions intimately related to the corresponding Sturm-Liouville
operator led in \cite{CKM}, \cite{CKT} and \cite{KT} to a deeper understanding
of the transmutation operators \cite{Gilbert}, \cite{Carroll} also known as
transformation operators \cite{Levitan}, \cite{Marchenko}. Indeed, the
functions $\varphi_{n}(x)$ resulted to be the images of the powers $x^{n}$
under the action of a corresponding transmutation operator \cite{CKT}. This
makes it possible to apply the transmutation operator even when the operator
itself is unknown (and this is the usual situation - there are very few
explicitly constructed examples available) due to the fact that its action on
any polynomial is known. This result was used in \cite{CKM} and \cite{CKT}
\ to prove the completeness (Runge-type approximation theorems) for families
of solutions of two-dimensional Schr\"{o}dinger and Dirac equations with
variable complex-valued coefficients.
\end{remark}

\begin{remark}
\label{RemSing}One of the functions $ru_{0}^{2}$ or $1/(pu_{0}^{2})$ may not
be continuous on $[a,b]$ and yet\/ $u_{1}$ or $u_{2}$ may make sense. For
example, in the case of the Bessel equation $(xu^{\prime})^{\prime}-\frac
{1}{x}u=-\lambda xu,$ we can choose $u_{0}(x)=x/2.$ Then $1/(pu_{0}^{2})\notin
C[0,1]$. Nevertheless, all integrals (\ref{Xtil}) exist and\/ $u_{1}$ coincides
with the nonsingular $\left(  1/\sqrt{\lambda}\right)  J_{1}(\sqrt{\lambda}%
x)$, while $u_{2}$ is a singular solution of the Bessel equation.
\end{remark}

\begin{remark}
\label{Rem Sturm separation theorem}When $p$ and $q$ are real-valued,
$p(x)\neq0$ for all $x\in\lbrack a,b]$ and $p$, $p^{\prime}$, $q$ are
continuous functions on $[a,b]$, the equation
\begin{equation}
Lv=0 \label{SL00}%
\end{equation}
is a regular Sturm-Liouville equation and possesses two linearly independent
real valued solutions $v_{1}$ and $v_{2}$. Due to Sturm's separation theorem
(see, e.g., \cite[p. 10]{King}) their zeros occur alternately and hence
$u_{0}=v_{1}+iv_{2}$ can be chosen as the required in theorem \ref{ThSolSL}
particular solution. If $r$ is a continuous on $[a,b]$ (in general,
complex-valued) function then the conditions of theorem \ref{ThSolSL} are fulfilled.
\end{remark}

The solutions $v_{1}$ and $v_{2}$ from Remark 4 can be in fact calculated using the same
procedure from Theorem \ref{ThSolSL}. Indeed, consider equation (\ref{SL00})
which can be written in the form%
\[
(pv^{\prime})^{\prime}=-qv.
\]
It has the form (\ref{SL}) with $r:=-q$, $\lambda=1$ and with a convenient
solution $v_{0}\equiv1$ of the (homogeneous) equation $(pv_{0}^{\prime
})^{\prime}=0$. Application of theorem \ref{ThSolSL} gives us the following
two linearly independent solutions of (\ref{SL00}),%
\begin{equation}
v_{1}=%
{\displaystyle\sum\limits_{k=0}^{\infty}}
\widetilde{Y}^{(2k)}\quad\text{and}\quad v_{2}=%
{\displaystyle\sum\limits_{n=0}^{\infty}}
Y^{(2k+1)} \label{v1v2}%
\end{equation}
where
\begin{equation}
\widetilde{Y}^{(0)}\equiv1,\quad Y^{(0)}\equiv1, \label{Y0}%
\end{equation}%
\begin{equation}
\widetilde{Y}^{(n)}(x)=\left\{
\begin{tabular}
[c]{ll}%
$-%
{\displaystyle\int\limits_{x_{0}}^{x}}
\widetilde{Y}^{(n-1)}(s)q(s)\,ds$, & $n$ \text{odd,}\\
$%
{\displaystyle\int\limits_{x_{0}}^{x}}
\widetilde{Y}^{(n-1)}(s)\frac{1}{p(s)}\,ds$, & $n$ \text{even,}%
\end{tabular}
\ \ \ \ \ \ \ \right.  \label{Ytil}%
\end{equation}%
\begin{equation}
Y^{(n)}(x)=\left\{
\begin{tabular}
[c]{ll}%
$%
{\displaystyle\int\limits_{x_{0}}^{x}}
Y^{(n-1)}(s)\frac{1}{p(s)}\,ds$, & $n$ \text{odd,}\\
$-%
{\displaystyle\int\limits_{x_{0}}^{x}}
Y^{(n-1)}(s)q(s)\,ds$, & $n$ \text{even},
\end{tabular}
\ \ \ \ \ \ \ \ \ \ \right.  \label{Y}%
\end{equation}
and the series in the equalities for $v_{1}$ and $v_{2}$ converge uniformly on
$[a,b]$.

Note that
\[
v_{1}(x_{0})=1,\quad v_{1}^{\prime}(x_{0})=0,\quad v_{2}(x_{0})=0,\quad
v_{2}^{\prime}(x_{0})=1/p(x_{0}).
\]
Solutions of (\ref{SL00}) in the form (\ref{v1v2}) is a long known result
(see, e.g., \cite{Weyl}).


Before we proceed to discuss eigenvalue and scattering problems it is worth
noticing that the representation of a general solution of the Sturm-Liouville
equation in the form of a spectral parameter power series (SPPS) given by
theorem \ref{ThSolSL} represents a natural and highly competitive method for
numerical solution of initial and boundary value problems. Compared to the
best standard routines it performs better or equal and with minimal
programmer's efforts. Moreover, the advantages of using SPPS become even more
transparent when the solution of the problem is required for many different
values of the spectral parameter. In such case the auxiliary functions
$X^{(n)}$ and $\widetilde{X}^{(n)}$, $n=0,1,2,\ldots$ should be computed only
once and then substitution of values of $\lambda$ into the expressions
(\ref{gensol}) gives us a solution of equation (\ref{SL}) for as many
different values of the spectral parameter as needed at no additional
computational cost. Nevertheless, first, let us show how SPPS performs at the
terrain of numerical ODE solvers for solution of initial value problems. In
\cite{CKKO2009} we made use of Matlab 7 and as a first step compared our
results with standard Matlab ODE solvers \cite{Ashino}, \cite{Shampine},
especially with ode45 which in the considered examples gave always better
results than other similar programs. Here we give two examples from
\cite{CKKO2009}.

For numerical approximations we consider partial sums of the infinite series
(\ref{gensol}) and (\ref{v1v2}), e.g., $u_{1}=u_{0}{\displaystyle\sum
\limits_{k=0}^{N}}\lambda^{k}\widetilde{X}^{(2k)}$ and $u_{2}=u_{0}%
{\displaystyle\sum\limits_{k=0}^{N}}\lambda^{k}X^{(2k+1)}$. \ The algorithm
was implemented in MATLAB. For the recursive integration we have chosen the
following strategy. On each step the integrand is represented through a cubic
spline using the spapi routine and the integration is performed using the
fnint routine (both from the spline toolbox of MATLAB).

Consider the following initial value problem for (\ref{SL00}): $p\equiv1$,
$q\equiv c^{2}$, $v(0)=1$, $v^{\prime}(0)=-1$ on the interval $(0,1)$. For
$c=1$ the absolute error of the result calculated by ode45 (with an optimal
tolerance chosen) was of order $10^{-9}$ and the relative error was of order
$10^{-6}$ whereas the absolute error of the result calculated with the aid of
the SPPS representation with $N$ from $55$ to $58$ was of order $10^{-16}$ and
the relative error was of order $10^{-14}$. Taking $c=10$ under the same
conditions the absolute and the relative errors of ode45 were of order
$10^{-6}$ and $10^{-5}$ respectively meanwhile our algorithm gave values of
order $10^{-12}$ in both cases. For the initial value problem: $p\equiv-1$,
$q\equiv c^{2}$, $v(0)=1$, $v^{\prime}(0)=-1$ on the interval $(0,1)$ in the
case $c=1$ the absolute and the relative errors of ode45 were of order
$10^{-8}$ whereas in our method this value was of order $10^{-15}$ already for
$N=50$. For $c=10$ the absolute and the relative errors of ode45 were of order
$10^{-3}$ and $10^{-7}$ respectively and in the case of our method these
values were of order $10^{-11}$ and $10^{-14}$ for $N=50$.

Consider yet another example. Let $p\equiv-1$, $q(x)=c^{2}x^{2}+c$ \ in
(\ref{SL00}). In this case the general solution has the form
\[
v(x)=e^{cx^{2}/2}\left(  c_{1}+c_{2}\int_{0}^{x}e^{-ct^{2}}dt\right)  .
\]
Take the same initial conditions as before, $v(0)=1$, $v^{\prime}(0)=-1$.
Then, while for $c=1$ the absolute and the relative error of ode45 were both
of order $10^{-8}$ and for $c=30$ the absolute error was $0.28$ and the
relative error was of order $10^{-6}$, our algorithm ($N=58$) gave the
absolute and relative errors of order $10^{-15}$ for $c=1$ and the absolute
and relative errors of order $10^{-9}$ and $10^{-15}$ respectively for $c=30.$
All calculations were performed on a common PC with the aid of Matlab 7.

The results of our numerical experiments show that in fact the SPPS
representations offer a powerful method for numerical solution of initial
value and boundary value problems for linear ordinary differential
second-order equations. The numerical calculation of the involved integrals
does not represent any considerable difficulty and can be done with a
remarkable accuracy.

Another observation which can be of use in many different situations is that
if for some purposes a derivative of the solution of (\ref{SL}) is required
there is no need to apply to the obtained solution an algorithm for numerical
differentiation. Instead, it is easy to see that%
\begin{equation}
u_{1}^{\prime}=\frac{u_{0}^{\prime}}{u_{0}}u_{1}+\frac{1}{u_{0}p}%
{\displaystyle\sum\limits_{k=1}^{\infty}}
\lambda^{k}\widetilde{X}^{(2k-1)}\quad\text{and\quad}u_{2}^{\prime}%
=\frac{u_{0}^{\prime}}{u_{0}}u_{2}+\frac{1}{u_{0}p}%
{\displaystyle\sum\limits_{k=0}^{\infty}}
\lambda^{k}X^{(2k)}. \label{u1u2prime}%
\end{equation}
Thus, the calculated auxiliary functions $X^{(n)}$ and $\widetilde{X}^{(n)}$,
$n=0,1,2,\ldots$ are used once more, this time for obtaining the derivative of
the solution.

\section{Solution of Sturm-Liouville problems\label{SectSturm-Liouville}}

In this section we outline the main ideas behind the application of the SPPS
method to the solution of Sturm-Liouville eigenvalue problems referring the
interested reader to \cite{KrPorter2010} for additional details and numerical
examples. The SPPS method allows one to reduce the Sturm-Liouville problem to
the problem of finding zeros of an analytic function of the complex variable
$\lambda$. Numerically the problem is reduced to finding roots of a polynomial
in $\lambda$. To find the precise expressions for Taylor coefficients of that
analytic function let us consider the general Sturm-Liouville problem with
unmixed boundary conditions. Thus, we look for the eigenvalues and
eigenfunctions of the problem
\begin{equation}
(pu^{\prime})^{\prime}+qu=\lambda ru, \label{SL1}%
\end{equation}%
\begin{equation}
u(a)\cos\alpha+u^{\prime}(a)\sin\alpha=0, \label{at0}%
\end{equation}%
\begin{equation}
u(b)\cos\beta+u^{\prime}(b)\sin\beta=0~, \label{atbeta}%
\end{equation}
where $[a,b]$ is a finite segment of the $x$-axis, $\alpha$ and $\beta$ are
arbitrary real numbers.

Let us choose the point $x_{0}$ from theorem \ref{ThSolSL} being equal to $a$
and consider the solutions $u_{1}$ and $u_{2}$ of (\ref{SL1}) defined by
(\ref{gensol}). Then from (\ref{at01}) and (\ref{at02}) we obtain that a
linear combination $u=c_{1}u_{1}+c_{2}u_{2}$ satisfies the following
conditions at $a$:%
\[
u(a)=c_{1}u_{0}(a)\quad\text{and\quad}u^{\prime}(a)=c_{1}u_{0}^{\prime
}(a)+c_{2}/(u_{0}(a)p(a)).
\]
Thus, in order that $u$ satisfy (\ref{at0}), the constants $c_{1}$ and $c_{2}$
must satisfy the equation
\[
c_{1}(u_{0}(a)\cos\alpha+u_{0}^{\prime}(a)\sin\alpha)+c_{2}\frac{\sin\alpha
}{u_{0}(a)p(a)}=0,
\]
which gives $c_{2}=\gamma c_{1}$ when $\alpha\neq\pi n$, with $\gamma
=-u_{0}(a)p(a)(u_{0}(a)\cot\alpha+u_{0}^{\prime}(a))$, whereas $c_{1}=0$ when
$\alpha=\pi n.$ In the latter case we have that if an eigenfunction of the
problem for a given $\lambda$ exists, up to a multiplicative constant it must
have a form $u=u_{2}$. The second boundary condition (\ref{atbeta}) together
with (\ref{u1u2prime}) leads to the following characteristic equation for the
eigenvalues%
\[
\cos\beta\,u_{0}(b)%
{\displaystyle\sum\limits_{k=0}^{\infty}}
\lambda^{k}X^{(2k+1)}(b)+\sin\beta\,\left(  u_{0}^{\prime}(b)%
{\displaystyle\sum\limits_{k=0}^{\infty}}
\lambda^{k}X^{(2k+1)}(b)+\frac{1}{u_{0}(b)p(b)}%
{\displaystyle\sum\limits_{k=0}^{\infty}}
\lambda^{k}X^{(2k)}(b)\right)  =0~,
\]
which is the same to
\begin{equation}%
{\displaystyle\sum\limits_{k=0}^{\infty}}
\lambda^{k}\left(  X^{(2k+1)}(b)\left(  \cos\beta\,u_{0}(b)+\sin\beta
\,u_{0}^{\prime}(b)\right)  +\frac{\sin\beta}{u_{0}(b)p(b)}X^{(2k)}(b)\right)
=0. \label{dispSL1}%
\end{equation}
Thus, the Sturm-Liouville problem (\ref{SL1})-(\ref{atbeta}) in the case
$\alpha=\pi n$ reduces to find zeros of the analytic function $%
{\displaystyle\sum\limits_{k=0}^{\infty}}
a_{k}\lambda^{k}$ where the Taylor coefficients $a_{k}$ have the form
$a_{k}=X^{(2k+1)}(b)\left(  \cos\beta\,u_{0}(b)+\sin\beta\,u_{0}^{\prime
}(b)\right)  +\frac{\sin\beta}{u_{0}(b)p(b)}X^{(2k)}(b)$.

Now let us suppose $\alpha\neq\pi n$. Then the boundary condition
(\ref{atbeta}) implies that%
\[
\left(  u_{0}(b)\cos\beta+u_{0}^{\prime}(b)\sin\beta\right)  \left(
{\displaystyle\sum\limits_{k=0}^{\infty}}
\lambda^{k}\widetilde{X}^{(2k)}(b)+\gamma%
{\displaystyle\sum\limits_{k=0}^{\infty}}
\lambda^{k}X^{(2k+1)}(b)\right)
\]%
\begin{equation}
+\frac{\sin\beta}{u_{0}(b)p(b)}\left(
{\displaystyle\sum\limits_{k=1}^{\infty}}
\lambda^{k}\widetilde{X}^{(2k-1)}(b)+\gamma%
{\displaystyle\sum\limits_{k=0}^{\infty}}
\lambda^{k}X^{(2k)}(b)\right)  =0. \label{dispSL2}%
\end{equation}
Thus the spectral problem (\ref{SL}), (\ref{at0}), (\ref{atbeta}) reduces to
the problem of calculating zeros of the analytic function $\kappa(\lambda)=%
{\displaystyle\sum\limits_{m=0}^{\infty}}
a_{m}\lambda^{m}$ where
\[
a_{0}=\left(  u_{0}(b)\cos\beta+u_{0}^{\prime}(b)\sin\beta\right)  (1+\gamma
X^{(1)}(b))+\frac{\gamma\sin\beta}{u_{0}(b)p(b)}%
\]
and%
\[
a_{m}=\left(  u_{0}(b)\cos\beta+u_{0}^{\prime}(b)\sin\beta\right)  \left(
\widetilde{X}^{(2m)}(b)+\gamma X^{(2m+1)}(b)\right)
\]%
\[
+\frac{\sin\beta}{u_{0}(b)p(b)}\left(  \widetilde{X}^{(2m-1)}(b)+\gamma
X^{(2m)}(b)\right)  ,\quad m=1,2,\ldots.
\]
This reduction of a Sturm-Liouville spectral problem to finding zeros of an
analytic function given by its Taylor series lends itself to a simple
numerical implementation. To calculate the first $n$ eigenvalues we consider
the Taylor polynomial $\kappa_{N}(\lambda)=%
{\displaystyle\sum\limits_{m=0}^{N}}
a_{m}\lambda^{m}$ with $N\geq n$. Thus the numerical approximation of
eigenvalues of the Sturm-Liouville problem reduces to the calculation of roots
of the polynomial $\kappa_{N}(\lambda)$.

In many physical applications (see \cite{BenAmara,Chanane2008,CodeBrowne2005,CoskunBayram2005, Fulton77,Walter}
and references therein) the Sturm-Liouville problems with boundary conditions
dependent on the spectral parameter arise. In this case together with equation
(\ref{SL1}) and boundary condition (\ref{at0}) the eigenfunction must satisfy
a second boundary condition of the form%

\begin{equation}
\beta_{1}u(b)-\beta_{2}u^{\prime}(b)=\varphi(\lambda)\left(  \beta_{1}%
^{\prime}u(b)-\beta_{2}^{\prime}u^{\prime}(b)\right)  , \label{SL53}%
\end{equation}
where $\varphi$ is a complex-valued function of the variable $\lambda$ and
$\beta_{1}$, $\beta_{2}$, $\beta_{1}^{\prime}$, $\beta_{2}^{\prime}$ are
complex numbers. For some special forms of the function $\varphi$ such as
$\varphi(\lambda)=\lambda$ or $\varphi(\lambda)=\lambda^{2}+c_{1}\lambda
+c_{2}$, results were obtained \cite{CodeBrowne2005}, \cite{Walter} concerning
the regularity of the problem (\ref{SL1}), (\ref{at0}), (\ref{SL53}); we will
not dwell upon the details.\ In general, the presence of the spectral
parameter in boundary conditions introduces additional considerable
difficulties both in theoretical and numerical analysis of the problems.
Nevertheless the SPPS approach gives a simple and natural insight into the
problem, and its practical application for numerical calculations is not in
fact more difficult than in the previously considered situation of $\lambda
$-independent boundary conditions.

For simplicity, let us suppose that $\alpha=0$ and hence the condition
(\ref{at0}) becomes $u(a)=0$. Then as was shown above, if an eigenfunction
exists it necessarily coincides with $u_{2}$ up to a multiplicative constant.
In this case condition (\ref{SL53}) becomes equivalent to the equality
\cite{KrPorter2010}%
\begin{equation}
\left(  u_{0}(b)\varphi_{1}(\lambda)-u_{0}^{\prime}(b)\varphi_{2}%
(\lambda)\right)
{\displaystyle\sum\limits_{k=0}^{\infty}}
\lambda^{k}X^{(2k+1)}(b)-\frac{\varphi_{2}(\lambda)}{u_{0}(b)p(b)}%
{\displaystyle\sum\limits_{k=0}^{\infty}}
\lambda^{k}X^{(2k)}(b)=0~, \label{eqSLparam}%
\end{equation}
where $\varphi_{1,2}(\lambda)=\beta_{1,2}-\beta_{1,2}^{\prime}\varphi
(\lambda)$. Calculation of eigenvalues given by (\ref{eqSLparam}) is
especially simple in the case of $\varphi$ being a polynomial of $\lambda$.
Precisely this particular situation was considered in all of the
above-mentioned references concerning Sturm-Liouville problems with spectral
parameter dependent boundary conditions. In any case the knowledge of an
explicit characteristic equation (\ref{eqSLparam}) for the spectral problem
(\ref{SL1}), (\ref{at0}), (\ref{SL53}) makes possible its accurate and
efficient solution.

The paper \cite{KrPorter2010} contains several numerical tests corresponding
to a variety of computationally difficult problems. All they reveal an
excellent performance of the SPPS method. We do not review them here referring
the interested reader to \cite{KrPorter2010}. Instead we consider another
interesting example, a Sturm-Liouville problem admitting complex eigenvalues.

Consider the equation (\ref{SL}) with $p\equiv-1$, $q\equiv0$ and $r\equiv1$
on the interval $(0,\pi)$ with the boundary conditions $u(0)=0$ and
$u(\pi)=-\lambda^{2}u(\pi)$. The exact eigenvalues of the problem are
$\lambda_{n}=n^{2}$ together with the purely imaginary numbers $\lambda_{\pm
}=\pm i$. Application of the SPPS method with $N=100$ and $3000$ interpolating
points (used for representing the integrands as splines) delivered the
following results $\lambda_{1}=1$, $\lambda_{2}=4.0000000000007$, $\lambda
_{3}=9.00000000001$, $\lambda_{4}=15.99999999996$, $\lambda_{5}=25.000000002$,
$\lambda_{6}=35.99999997$, $\lambda_{7}=49.0000004$, $\lambda_{8}=63.9999994$,
$\lambda_{9}=80.9996$, $\lambda_{10}=100.02$ and $\lambda_{\pm}=\pm i$. Thus,
the complex eigenvalues are as easily and accurately detected by the SPPS
method as the real eigenvalues. Note that for a better accuracy in calculation
of higher eigenvalues of a Sturm-Liouville problem an additional simple
shifting procedure (described in \cite{KrPorter2010}) based on the
representation of solutions of (\ref{SL}) not as a series in powers of
$\lambda$ but in powers of $(\lambda-\lambda_{0})$ is helpful. We did not
apply it here and hence the accuracy of the calculated value of $\lambda_{10}$
is considerably worse than the accuracy of the first calculated eigenvalues
which in general can be improved by means of the mentioned shifting procedure.

\section{Periodic potentials: Floquet-Bloch solutions and Hill's discriminant
\label{SectPeriodic}}

Towards the end of the 19th century, Hill \cite{Hill86} and Floquet
\cite{Floquet83} initiated the rigorous study of the spectral properties of
periodic Sturm-Liouville equations with real coefficients and with periodic
(and antiperiodic) conditions imposed to their solutions $\mathbf{y}%
(0)=\pm\mathbf{y}(\pi)$, where $\mathbf{y}=(y,y^{\prime})^{Tr}$. In 1883,
Floquet established Floquet's theorem asserting that every solution is a
linear superposition of independent solutions, both of the form of exponential
factors multiplied by periodic functions, while Hill's work published in 1877
in Cambridge, Mass., and reprinted in Europe in 1886, deals with a special
component of the motion of the lunar perigee. Other fundamental results
concerning the sequence of eigenvalues have been obtained by Lyapunov in 1902
\cite{Liapunov}.

The paper of Hill made this class of equations of interest for many authors
and the term Hill equations has been commonly used since about a century for
second-order linear differential equations with periodic coefficients, in
particular for $y^{\prime\prime}+f(t)y=0$ with $f(t)$ a periodic function.
There are three basic methods to determine the existence of eigenvalues in
this case, (i) through Floquet theory and Hill's discriminant $D(\lambda)$,
(ii) using operator theory and variational techniques, and (iii) via
Pr\"ufer's transformation. The standard method is the first one which
essentially reduces itself to seeking the nontrivial solutions (also known as
quasi-periodic solutions) of a Hill-type equation that satisfy
\[
\mathbf{y}(\pi)=\beta\mathbf{y}(0)
\]
for some complex parameters $\beta$ known as `Floquet multipliers' which solve
the quadratic equation
\[
\beta^{2}-D(\lambda)\beta+1=0~.
\]
Thus, to get the Floquet multipliers one needs Hill's discriminant, which in
general is a real-valued function expressed in terms of fundamental solutions
of the given Hill equation. Notice that from the standpoint of the Floquet
multipliers the periodic and antiperiodic solutions are special cases
corresponding to $\beta_{p}=e^{2k\pi\, i}$ and $\beta_{a}=e^{(2k+1)\pi\, i}$,
respectively, for any integer $k$.

In the realm of quantum mechanics, the first fundamental application of
Floquet theory belongs to Bloch \cite{Bloch} in 1928, who, using the quantum
physical terminology, introduced Floquet's theorem for the special case in
which the exponential factors are plane waves in the important context of
solid state physics. Interestingly, no reference to the mathematical
literature is given in Bloch's paper. In 1931, Kr\"onig and Penney \cite{KPen}
introduced the first model of bands (stability regions) and gaps (instability
regions) for the motion of electrons in crystal lattices. The activity in this
area has been retaken only after the second world war period
\cite{James,Case,Slater,Scarf} and at the present time a vigorous progress
takes place in extended research lines covering nanostructures and photonic crystals.

In the rest of this section, we briefly describe our recent result
\cite{KiraRosu2010} of expressing the Hill discriminant in terms of SPPS for a
Hill-Sturm-Liouville equation of the type:
\begin{equation}
-(p(x)f^{\prime}(x,\lambda))^{\prime}+q(x)f(x,\lambda)=\lambda f(x,\lambda)~.
\label{eq}%
\end{equation}
We assume that $p(x)>0$, $p^{\prime}(x)$ and $q(x)$ are continuous bounded
periodic functions of period $T$.

We first recall some necessary definitions and basic properties
from the Floquet (Bloch) theory. For more details see, e.g., \cite{Magnus,
Eastham}.

For each $\lambda$ there exists a fundamental system of solutions, i.e., two
linearly independent solutions of (\ref{eq}) $f_{1}(x,\lambda)$ and
$f_{2}(x,\lambda)$ which satisfy the initial conditions%
\begin{equation}
f_{1}(0,\lambda)=1,\quad f_{1}^{\prime}(0,\lambda)=0,\quad f_{2}%
(0,\lambda)=0,\quad f_{2}^{\prime}(0,\lambda)=1. \label{cond}%
\end{equation}
The Hill discriminant has the following expression in terms of the fundamental
system of solutions
\[
D(\lambda)=f_{1}(T,\lambda)+f_{2}^{\prime}(T,\lambda).
\]
Employing $D(\lambda)$ one can easily describe the spectrum of the
corresponding equation. Namely, the values of $\lambda$ for which $\left\vert
D(\lambda)\right\vert \leq2$ form the allowed bands or stability intervals
meanwhile the values of $\lambda$ such that $\left\vert D(\lambda)\right\vert
>2$ belong to forbidden bands or instability intervals \cite{Magnus}. The band
edges (values of $\lambda$ such that $\left\vert D(\lambda)\right\vert =2$)
represent the discrete spectrum of the operator, i.e., they are the
eigenvalues of the operator with periodic ($D(\lambda)=2$) or antiperiodic
($D(\lambda)=-2$) boundary conditions.
The eigenvalues $\lambda_{n}$, $n=0,1,2,...$ form an infinite sequence
$\lambda_{0}<\lambda_{1}\leqslant\lambda_{2}<\lambda_{3}...$, and an important
property of the minimal eigenvalue $\lambda_{0}$ is the existence of a
corresponding periodic nodeless solution $f_{0}(x,\lambda_{0})$ \cite{Magnus}.
In general solutions of (\ref{eq}) are not of course periodic, and one of the
important tasks related to Sturm-Liouville equations with periodic
coefficients is the construction of quasiperiodic solutions. In this paper, we
use the matching procedure from \cite{James} for which the main ingredient is
the pair of solutions $f_{1}(x,\lambda)$ and $f_{2}(x,\lambda)$ of (\ref{eq})
satisfying conditions (\ref{cond}). Namely, using $f_{1}(x,\lambda)$ and
$f_{2}(x,\lambda)$ one obtains the quasiperiodic solutions $f_{\pm}%
(x+T)=\beta_{\pm}f_{\pm}(x)$ as follows
\begin{equation}
f_{\pm}(x,\lambda)=\beta_{\pm}^{n}F_{\pm}(x-nT,\lambda),\quad\left\{
\begin{array}
[c]{c}%
nT\leq x<(n+1)T\\
n=0,\pm1,\pm2,...~
\end{array}
\right.  , \label{bloch}%
\end{equation}
where $F_{\pm}(x,\lambda)$ are the so-called self-matching solutions, which
are the following linear combinations\ $F_{\pm}(x,\lambda)=f_{1}%
(x,\lambda)+\alpha_{\pm}f_{2}(x,\lambda)$\ with $\alpha_{\pm}$ being roots of
the algebraic equation $f_{2}(T,\lambda)\alpha^{2}+(f_{1}(T,\lambda
)-f_{2}^{\prime}(T,\lambda))\alpha-f_{1}^{\prime}(T,\lambda)=0$. The Bloch
factors $\beta_{\pm}$ are a measure of the rate of increase (or decrease) in
magnitude of the self-matching solutions $F_{\pm}(x,\lambda)$ when one goes
from the left end of the cell to the right end, i.e., $\beta_{\pm}%
(\lambda)=\frac{F_{\pm}(T,\lambda)}{F_{\pm}(0,\lambda)}$. The values of
$\beta_{\pm}$ are directly related to the Hill discriminant, $\beta_{\pm
}(\lambda)=\frac{1}{2}(D(\lambda)\mp\sqrt{D^{2}(\lambda)-4})$, and obviously
at the band edges $\beta_{+}=\beta_{-}=\pm1$ for $D(\lambda)=\pm2$, correspondingly.

\subsection{The SPPS series representation of Hill's discriminant}


The SPPS construction method of the solutions $f_{1}(x,\lambda)$ and
$f_{2}(x,\lambda)$
satisfying the initial conditions (\ref{cond}) is based on the knowledge of
one non-vanishing particular solution $f_{0}(x,\lambda_{0})$
bounded on $[0,T]$ together with $\frac{1}{f_{0}(x,\lambda_{0})}$. In the case
of Hill's equation the first eigenvalue $\lambda_{0}$
generates a nodeless periodic eigenfunction $f_{0}(x,\lambda_{0})$. In what
follows, we\ initially suppose that the value of $\lambda_{0}$ is known. Note
that, it can be obtained by different methods including the same SPPS method
\cite{KrPorter2010} as we explain in subsection \ref{3.4}.

Given $\lambda_{0}$, we proceed in three steps in order to obtain the
representation of Hill's discriminant:

\begin{quote}

\noindent$\bullet$ the first one is the construction of a particular nodeless
solution $f_{0}(x,\lambda_{0})$
which is periodic, i.e., $f_{0}(x+T,\lambda_{0})=f_{0}(x,\lambda_{0})$,

\noindent$\bullet$ the second one is the construction of the fundamental
system of solutions $f_{1}(x,\lambda)$ and $f_{2}(x,\lambda)$
for all values of the parameter $\lambda$,

\noindent$\bullet$ the final step is getting the representation of Hill's discriminant.

\end{quote}

We detail each of the steps in the following subsections.


\subsection{The nodeless periodic solution}

\label{3.1}

We want to obtain the nodeless periodic solution $f_{0}(x,\lambda_{0})$
for $\lambda=\lambda_{0}$ of the equation
\begin{equation}
-(p(x)(f_{0}(x))^{\prime})^{\prime}+q(x)f_{0}(x)=\lambda_{0}f_{0}(x)~.
\label{lambda0}%
\end{equation}
To achieve this goal, we first have to construct in SPPS form the fundamental
system of solutions of (\ref{lambda0}).
These solutions are not necessarily periodic. However, one can follow the old
procedure of James \cite{James} allowing to obtain from $f_{0,1}(x,\lambda
_{0})$ and $f_{0,2}(x,\lambda_{0})$ the Floquet type solutions which
degenerate to a single periodic/antiperiodic solution $f_{0}(x,\lambda_{0})$
since $\lambda_{0}$ represents a band edge.

The functions $f_{0,1}(x,\lambda_{0})$ and $f_{0,2}(x,\lambda_{0})$ can be
calculated according to iteration formulas of the type (\ref{Y0})-(\ref{Y})
\begin{equation}
f_{0,1}(x,\lambda_{0})= {\displaystyle\sum\limits_{even\hspace{0.05in}%
n=0}^{\infty}} \,\widetilde{X}_{0}^{(n)}\qquad\text{and}\,\qquad
f_{0,2}(x,\lambda_{0})=p(0) {\displaystyle\sum\limits_{odd\hspace{0.05in}%
n=1}^{\infty}} X_{0}^{(n)}, \label{f01 f02}%
\end{equation}
where
\[
\widetilde{X}_{0}^{(0)}\equiv1,\qquad X_{0}^{(0)}\equiv1,
\]
\[
\widetilde{X}_{0}^{(n)}(x)=\left\{
\begin{tabular}
[c]{ll}%
$\int_{0}^{x}\widetilde{X}_{0}^{(n-1)}(\xi)(q(\xi)-\lambda_{0})d\xi\qquad$ &
$\text{for an odd }n$\\
& \\
$\int_{0}^{x}\widetilde{X}_{0}^{(n-1)}(\xi)\frac{1}{p(\xi)}d\xi\qquad$ &
$\text{for an even }n$%
\end{tabular}
\right.
\]

\[
X_{0}^{(n)}(x)=\left\{
\begin{tabular}
[c]{ll}%
$\int_{0}^{x}X_{0}^{(n-1)}(\xi)\frac{1}{p(\xi)}d\xi\qquad$ & $\text{for an odd
}n$\\
& \\
$\int_{0}^{x}X_{0}^{(n-1)}(\xi)(q(\xi)-\lambda_{0})d\xi\qquad$ & $\text{for an
even }n$~.
\end{tabular}
\ \right.
\]
The periodic nodeless solution of (\ref{lambda0}) is constructed as a
particular case of a quasi-periodic solution (\ref{bloch}), essentially as a
self-matching solution, i.e.,
\begin{align}
f_{0}(x,\lambda_{0})  &  =f_{0,1}(x-nT,\lambda_{0})+\alpha_{p} f_{0,2}%
(x-nT,\lambda_{0}),\label{f0}\\
&  \left\{
\begin{array}
[c]{c}%
nT\leq x<(n+1)T\\
n=0,1,2,...~,
\end{array}
\right. \nonumber
\end{align}
since the Floquet phase multiplier is $\beta=1$ in the periodic case and
$\alpha_{p}=\frac{f_{0,2}^{\prime}(T,\lambda_{0})-f_{0,1}(T,\lambda_{0}%
)}{2f_{0,2}(T,\lambda_{0})}$, see \cite{James}.

\subsection{Fundamental system of solutions}

\label{3.2}

Once having the function $f_{0}(x,\lambda_{0})$, the solutions $f_{1}%
(x,\lambda)$ and $f_{2}(x,\lambda)$
for all values of the parameter $\lambda$ can be given using the SPPS method
once again
\begin{align}
f_{1}(x,\lambda)  &  =\frac{f_{0}(x)}{f_{0}(0)}\widetilde{\Sigma}%
_{0}(x,\lambda,\lambda_{0})+p(0)f_{0}^{\prime}(0)f_{0}(x)\Sigma_{1}%
(x,\lambda,\lambda_{0}),\nonumber\\
& \label{f1 f2}\\
f_{2}(x,\lambda)  &  =-p(0)f_{0}(0)f_{0}(x)\Sigma_{1}(x,\lambda,\lambda
_{0}).\nonumber
\end{align}
The SPPS summations $\widetilde{\Sigma}_{0}$ and $\Sigma_{1}$ have the following expressions
\[
\widetilde{\Sigma}_{0}(x,\lambda,\lambda_{0})=\sum_{\,n=0}^{\infty}%
\widetilde{X}^{(2n)}(x)(\lambda-\lambda_{0})^{n},\quad\Sigma_{1}%
(x,\lambda,\lambda_{0})=\sum_{n=1}^{\infty}X^{(2n-1)}(x)(\lambda-\lambda
_{0})^{n-1}~,
\]
where the coefficients $\widetilde{X}^{(n)}(x)$, $X^{(n)}(x)$ are given by the
recursive relations
\[
\widetilde{X}^{(0)}\equiv1,\qquad X^{(0)}\equiv1,
\]
%

\begin{equation}
\tilde{X}^{(n)}(x)=
\begin{cases}
\int_{0}^{x}\tilde{X}^{(n-1)}(\xi)f_{0}^{2}(\xi)d\xi\qquad\mathrm{for}%
\,\mathrm{an}\,\mathrm{odd}\,n\\
\\
-\int_{0}^{x}\tilde{X}^{(n-1)}(\xi)\frac{d\xi}{p(\xi)f_{0}^{2}(\xi)}%
\qquad\ \ \ \ \mathrm{for}\,\mathrm{an}\,\mathrm{even}\,n
\end{cases}
\label{K1}%
\end{equation}

\bigskip%

\begin{equation}
X^{(n)}(x)=
\begin{cases}
-\int_{0}^{x}X^{(n-1)}(\xi)\frac{d\xi}{p(\xi)f_{0}^{2}(\xi)}\qquad
\ \ \ \ \ \mathrm{for}\,\mathrm{an}\,\mathrm{odd}\,n\\
\\
\int_{0}^{x}X^{(n-1)}(\xi)f_{0}^{2}(\xi)d\xi\qquad\ \mathrm{for}
\,\mathrm{an}\,\mathrm{even}\,n~,
\end{cases}
\label{K2}%
\end{equation}
which are identical to Eqs.~(\ref{X0})-(\ref{X}) unless for obvious sign changes.

One can check by a straightforward calculation that the solutions $f_{1}$ and
$f_{2}$ fulfill the initial conditions (\ref{cond}).
Having obtained the fundamental system of solutions for any value of $\lambda
$, one can apply the construction (\ref{bloch}) in order to obtain the Bloch
solutions which become eigenfunctions for $\lambda$ being eigenvalues.

\subsection{\bigskip Hill's discriminant in SPPS form}

\label{3.3}

We are ready now to write the Hill discriminant $D(\lambda)=f_{1}%
(T,\lambda)+f_{2}^{\prime}(T,\lambda)$ in a simple explicit form using the
SPPS expressions of $f_{1}(T,\lambda)$ and $f_{2}^{\prime}(T,\lambda)$ in
(\ref{f1 f2})
\begin{align}
D(\lambda)  &  =\frac{f_{0}(T)}{f_{0}(0)}\widetilde{\Sigma}_{0}(T,\lambda
,\lambda_{0})+\frac{f_{0}(0)}{f_{0}(T)}\Sigma_{0}(T,\lambda,\lambda
_{0})\label{DD}\\
&  +\left(  f_{0}^{\prime}(0)f_{0}(T)-f_{0}(0)f_{0}^{\prime}(T)\right)
p(0)\Sigma_{1}(T,\lambda,\lambda_{0})~.\nonumber
\end{align}
Finally, taking into account that $f_{0}(x)$ is a $T$-periodic function
$f_{0}(0)=f_{0}(T)$ and\ writing the explicit expressions for $\widetilde
{\Sigma}_{0}(T,\lambda,\lambda_{0})$ and $\Sigma_{0}(T,\lambda,\lambda_{0})$
we obtain a representation for Hill's discriminant associated with (\ref{eq})
\begin{equation}
D(\lambda)\equiv\sum_{n=0}^{\infty}\left(  \tilde{X}^{(2n)}(T)+X^{(2n)}%
(T)\right)  (\lambda-\lambda_{0})^{n}\text{.} \label{D}%
\end{equation}
Thus, only one particular nodeless and periodic solution $f_{0}(x,\lambda
_{0})$ of (\ref{eq}) is needed for constructing the associated Hill
discriminant. We formulate the result (\ref{D}) as the following theorem:

\begin{theorem}
Let $\lambda_{0}$ be the lowest eigenvalue of the periodic Sturm-Liouville
problem (\ref{eq}) on the segment $[0,T]$ with periodic boundary conditions
and $f_{0}(x,\lambda_{0})$ be the corresponding eigenfunction. Then the Hill
discriminant for (\ref{eq}) has the form (\ref{D}) where $\tilde{X}^{(2n)}$
and $X^{(2n)}$ are calculated according to (\ref{K1}) and (\ref{K2}), and the
series converges uniformly on any compact set of values of $\lambda$.
\end{theorem}

To illustrate the formula (\ref{D}) we consider a simple example. Let
$q(x)=0$, $p(x)=1$ in equation (\ref{eq}). It is easy to see that the
associated discriminant is $D(\lambda)=2\cos\sqrt{\lambda}T$, from where we
obtain $\lambda_{0}=0$\ and a corresponding non-trivial periodic solution is
$f_{0}(x)=1$. Now making use of this solution we construct the discriminant by
means of the formula (\ref{D}). The coefficients $\tilde{X}^{(2n)}(T)$ and
$X^{(2n)}(T)$ given by (\ref{K1}) and (\ref{K2}) take the form
\[
\tilde{X}^{(2n)}(T)=X^{(2n)}(T)=(-1)^{n}\frac{T^{2n}}{(2n)!},\quad
n=0,1,2,...~.
\]
The substitution in (\ref{D}) gives $D(\lambda)=2\cos\sqrt{\lambda}T$.

\subsection{Construction of the first eigenvalue $\lambda_{0}$ by the SPPS
method}

\label{3.4}

Notice that in the expression (\ref{DD}) for $D(\lambda)$ and in all
reasonings previous to it we do not use the periodicity of the solution
$f_{0}(x,\lambda_{0})$, therefore (\ref{DD}) and the whole procedure for
obtaining it are valid for any $\lambda_{\ast}$ such that there exists a
corresponding solution $f_{\ast}(x,\lambda_{\ast})$ which is bounded on
$[0,T]$\ together with $1/(pf_{\ast}^{2})$. Such a solution $f_{\ast
}(x,\lambda_{\ast})$ can be obtained in the following way
\begin{equation}
f_{\ast}(x,\lambda_{\ast})=f_{\ast,1}(x,\lambda_{\ast})+if_{\ast,2}%
(x,\lambda_{\ast}) \label{f_asterisc}%
\end{equation}
where $f_{\ast,1}(x,\lambda_{\ast})$ and $f_{\ast,2}(x,\lambda_{\ast})$ are
given by (\ref{f01 f02}) with $\lambda_{\ast}$ instead of $\lambda_{0}$. For
more details see \cite{KrCV2008}. The pair of the independent solutions
$f_{1}(x,\lambda)$ and $f_{2}(x,\lambda)$ of (\ref{eq}) given by (\ref{f1 f2})
of course are independent of the choice of the solution $f_{0}(x,\lambda_{0}%
)$, hence instead of $f_{0}(x,\lambda_{0})$ in (\ref{DD}) one can take
$f_{\ast}(x,\lambda_{\ast})$ given by (\ref{f_asterisc}). \ Thus, in terms of
$f_{\ast}(x,\lambda_{\ast})$ where $\lambda_{\ast}$ is essentially arbitrary,
$D(\lambda)$ can be represented as a series in powers of $(\lambda
-\lambda_{\ast})$%
\begin{align}
D(\lambda)  &  =\sum_{n=0}^{\infty}\left(  \frac{f_{\ast}(T)}{f_{\ast}%
(0)}\tilde{X}^{(2n)}(T)+\frac{f_{\ast}(0)}{f_{\ast}(T)}X^{(2n)}(T)+\right.
\label{Dstar}\\
&  +\left(  f_{\ast}^{\prime}(0)f_{\ast}(T)-f_{\ast}(0)f_{\ast}^{\prime
}(T)\right)  p(0)X^{(2n+1)}(T)%
\genfrac{.}{)}{0pt}{}{{}}{{}}
(\lambda-\lambda_{\ast})^{n}\text{.}\nonumber
\end{align}

Now the band edge $\lambda_{0}$ required for the formula (\ref{D}) can be
calculated as a first zero of the expression $D(\lambda)-2$ where $D(\lambda)$
is given by (\ref{Dstar}). For the numerical purpose it can be useful to know
the interval containing $\lambda_{0}$. Since $q$ is a bounded periodic
function, there is a number $\Lambda$ which satisfies the inequality
$q(x)>\Lambda$ $\forall x\in\mathbf{R}$. It is known \cite{Eastham} that
$D(\lambda)>2$ for all $\lambda\leqslant\Lambda$, therefore the lower estimate
for $\lambda_{0}$ is the following
\[
\lambda_{0}\geqslant\min q(x).
\]
The upper bound can be obtained considering the Rayleigh quotient for periodic
problems \cite{Pinchover}
\[
\lambda_{0}\leqslant\frac{\int_{0}^{T}\left(  p\left(  x\right)  (u^{\prime
}\left(  x\right)  )^{2}+q\left(  x\right)  \left(  u(x)\right)  ^{2}\right)
dx}{\int_{0}^{T}\left(  u(x)\right)  ^{2}dx}~,
\]
where $u(x)\in\mathbf{C}^{2}\left[  0,T\right]  $ is periodic with period $T$.
The equality occurs if and only if $u(x)$ is an eigenfunction corresponding to
$\lambda_{0}$.

\subsection{Hill's discriminant of the supersymmetric (SUSY)-related equation}

\label{ss4-6}

In this subsection, we consider the SUSY partner equation of Eq.~(\ref{eq})
and write down the SPPS form of its solutions. The latter allow us to prove
the equality between the Hill discriminants of equation (\ref{eq}) and its
SUSY-related Eq.~(\ref{eq1}). For various aspects of SUSY periodic problems,
see \cite{Samsonov,Correa,Cooper}.

The left-hand side of the equation (\ref{eq}) can be factorized in the
following way \cite{Plastino}
\begin{equation}
\left(  -d_{x}p^{\frac{1}{2}}(x)+\Phi(x)\right)  \left(  p^{\frac{1}{2}%
}(x)d_{x}+\Phi(x)\right)  f(x), \label{fact}%
\end{equation}
where $d_{x}$ means the $x$-derivative, the superpotential $\Phi(x)$ is
defined as follows $\Phi(x)=-p^{\frac{1}{2}}(x)\frac{f_{0}^{\prime}%
(x,\lambda_{0})}{f_{0}(x,\lambda_{0})}$. Using this factorization the
coefficient $q(x)$ can be expressed as
\[
q(x)=\Phi^{2}(x)-\left(  p^{\frac{1}{2}}(x)\Phi(x)\right)  ^{\prime}%
+\lambda_{0}.
\]
Introducing the following Darboux transformation
\begin{equation}
\left(  p^{\frac{1}{2}}(x)d_{x}+\Phi(x)\right)  f(x,\lambda)=\tilde
{f}(x,\lambda), \label{Darb}%
\end{equation}
one obtains the equation supersymmetrically related to equation (\ref{eq})
\[
\left(  p^{\frac{1}{2}}(x)d_{x}+\Phi(x)\right)  \left(  -d_{x}p^{\frac{1}{2}%
}(x)+\Phi(x)\right)  \tilde{f}(x,\lambda)=\lambda\tilde{f}(x,\lambda),
\]
which can be written as follows
\begin{equation}
-d_{x}(p(x)d_{x}\tilde{f}(x,\lambda))+\tilde{q}(x)\tilde{f}(x,\lambda
)=\lambda\tilde{f}(x,\lambda)~, \label{eq1}%
\end{equation}
where $\tilde{q}(x)$ is the SUSY partner of the coefficient $q(x)$ given by%

\begin{equation}
\tilde{q}(x)=q(x)+2p^{\frac{1}{2}}(x)\Phi^{\prime}(x)-p^{\frac{1}{2}%
}(x)(p^{\frac{1}{2}}(x))^{\prime\prime}~. \label{q2}%
\end{equation}
It is worth noting that as $\Phi(x)$ is a $T$-periodic function, the Darboux
transformation assures the $T$-periodicity of $\tilde{q}(x)$. In addition,
when $p(x)$ is a constant, the SL coefficient $q$ is a quantum-mechanical
potential, while $\tilde{q}(x)$ is its Darboux counterpart also termed a
supersymmetric partner in quantum mechanics.

The pair of linearly independent solutions $\tilde{f}_{1}(x,\lambda)$ and
$\tilde{f}_{2}(x,\lambda)$\ of (\ref{eq1}) can be obtained directly from the
solutions (\ref{f1 f2}) by means of the Darboux transformation (\ref{Darb}).
We additionally take the linear combinations in order that the solutions
$\tilde{f}_{1}(x,\lambda)$ and $\tilde{f}_{2}(x,\lambda)$\ satisfy the initial
conditions $\tilde{f}_{1}(0,\lambda)=\tilde{f}_{2}^{\prime}(0,\lambda)=1$ and
$\tilde{f}_{1}^{\prime}(0,\lambda)=\tilde{f}_{2}(0,\lambda)=0$
\begin{align}
\tilde{f}_{1}(x,\lambda)  &  =\frac{p^{\frac{1}{2}}(0)f_{0}(0)}{p^{\frac{1}%
{2}}(x)f_{0}(x)}\Sigma_{0}(x,\lambda)+\frac{[p^{\frac{1}{2}}(x)]^{\prime
}|_{x=0}-\Phi(0)}{\left(  \lambda-\lambda_{0}\right)  f_{0}(0)p^{\frac{1}{2}%
}(x)f_{0}(x)} \widetilde{\Sigma}_{1}(x,\lambda),\label{g1}\\
\tilde{f}_{2}(x,\lambda)  &  =\frac{p^{\frac{1}{2}}(0)}{\left(  \lambda
-\lambda_{0}\right)  f_{0}(0)p^{\frac{1}{2}}(x)f_{0}(x)}\widetilde{\Sigma}%
_{1}(x,\lambda). \label{g2}%
\end{align}

These two solutions allow us to write the expression for Hill's discriminant
associated to the equation (\ref{eq1}), that is $\widetilde{D}(\lambda
)=\tilde{f}_{1}(T,\lambda)+\tilde{f}_{2}^{\prime}(T,\lambda)$. This requires
the expression of the derivative
of $\tilde{f}_{2}(x,\lambda)$ and evaluating it for $x=T$.
In addition, one should notice that as the functions $f_{0}(x,\lambda_{0})$
and $p(x)$ are $T$-periodic, i.e., $f_{0}(0,\lambda_{0})=f_{0}(T,\lambda_{0})$
and $p(0)=p(T)$, then obviously, the functions $f_{0}^{\prime}(x,\lambda
_{0}),p^{\frac{1}{2}}(x)$ and $[p^{\frac{1}{2}}(x)]^{\prime}$ possess the same
properties. The result is \cite{KiraRosu2010}
\begin{align*}
\widetilde{D}(\lambda)  &  =\Sigma_{0}(T,\lambda)+\widetilde{\Sigma}%
_{0}(T,\lambda)+\left(  \frac{[p^{\frac{1}{2}}(x)]^{\prime}|_{x=0}-\Phi
(0)}{\left(  \lambda-\lambda_{0}\right)  f_{0}(0,\lambda_{0})p^{\frac{1}{2}%
}(T)f_{0}(T,\lambda_{0})}-\right. \\
&  \left.  -\frac{[p^{\frac{1}{2}}(x)]^{\prime}|_{x=T}f_{0}(T,\lambda
_{0})+p^{\frac{1}{2}}(T)f_{0}^{\prime}(T,\lambda_{0})}{\left(  \lambda
-\lambda_{0}\right)  f_{0}(0,\lambda_{0})p^{\frac{1}{2}}(T)f_{0}^{2}%
(T,\lambda_{0})}\right)  \widetilde{\Sigma}_{1}(T,\lambda).
\end{align*}
The substitution $\Phi(0)=-p^{\frac{1}{2}}(0)\frac{f_{0}^{\prime}%
(0,\lambda_{0})}{f_{0}(0,\lambda_{0})}$ clearly shows that the expression in
brackets vanishes leading to the simple formula
\[
\widetilde{D}(\lambda)=\Sigma_{0}(T,\lambda)+\widetilde{\Sigma}_{0}%
(T,\lambda)=\sum_{n=0}^{\infty}\left(  \tilde{X}^{(2n)}(T)+X^{(2n)}(T)\right)
(\lambda-\lambda_{0})^{n}~,
\]
which is identical to (\ref{D}) and therefore
\begin{equation}
D(\lambda)\equiv\widetilde{D}(\lambda). \label{DD1}%
\end{equation}

Thus, we can make the following statement:


\begin{theorem}
Let $\lambda_{0}$ be the first eigenvalue of (\ref{eq}) and $f_{0}%
(x,\lambda_{0})$ the corresponding $T$-periodic nodeless eigenfunction. Then
the Darboux transformation (\ref{Darb}) with $\Phi(x)=-p^{\frac{1}{2}}%
(x)\frac{f_{0}^{\prime}(x,\lambda_{0})}{f_{0}(x,\lambda_{0})}$ leads to a
SUSY-related Eq.~(\ref{eq1}) with the preservation of the Hill discriminant,
i.e., Eq.~(\ref{DD1}) holds.
\end{theorem}

From the identity of discriminants (\ref{DD1}) it is clear that $\lambda_{0} $
gives rise to a nodeless periodic solution $\tilde{f}_{0}(x,\lambda_{0})$ of
Eq.~(\ref{eq1}). Taking $\lambda=\lambda_{0}$ in (\ref{g1}) and (\ref{g2}) we
get this eigenfunction in the form $\tilde{f}_{0}(x,\lambda_{0})=\frac
{1}{p^{\frac{1}{2}}(x)f_{0}(x,\lambda_{0})}~.$

Notice that, the factorization method can be applied to Eq.~(\ref{eq1}) with
the superpotential $\Phi_{1}(x)=-p^{\frac{1}{2}}(x)\frac{\tilde{f}%
_{0}^{^{\prime}}(x,\lambda_{0})}{\tilde{f}_{0}(x,\lambda_{0})}$. In this case,
we obtain the representation
\[
\tilde{q}=\Phi_{1}^{2}(x)-\left(  p^{\frac{1}{2}}(x)\Phi_{1}(x)\right)
^{\prime}+\lambda_{0},
\]
which reduces to the equality (\ref{q2}) if one notices the relationship
$\Phi_{1}(x)=(p^{\frac{1}{2}}(x))^{\prime}-\Phi(x)$. It can be also shown that
$\tilde{\tilde q}\equiv q$, where $\tilde{\tilde q}=\tilde{q}(x)+2p^{\frac
{1}{2}}(x)\Phi_{1}^{\prime}(x)-p^{\frac{1}{2}}(x)(p^{\frac{1}{2}}%
(x))^{\prime\prime}$ is the superpartner potential of $\tilde{q}(x)$. Thus,
the Darboux transformation (\ref{Darb}) with the superpotential $\Phi_{1}(x)$
applied to Eq.~(\ref{eq1}) does not produce a different potential.

\subsection{Numerical calculation of the eigenvalues based on Hill's
discriminant in SPPS form}

\label{ss4-7}

As is well known, see e.g., \cite{Magnus}, the zeros of the functions
$D(\lambda)\pm2$ represent \ eigenvalues of the corresponding operator. In
this section, we show that besides other possible applications the
representation (\ref{D}) gives us an efficient tool for the calculation of the
discrete spectrum of a periodic Sturm-Liouville operator.

The first step of the numerical realization of the method consists in
calculation of the minimal eigenvalue $\lambda_{0}$ by means of the procedure
given in subsection \ref{3.4} and subsequently in construction of the
corresponding nodeless periodic solution $f_{0}(x,\lambda_{0})$ using formula
(\ref{f0}). The next step of the algorithm is to compute the functions
$\tilde{X}^{(n)}$ and $X^{(n)}$ given by (\ref{K1}) and (\ref{K2}),
respectively. This construction is based on the eigenfunction $f_{0}%
(x,\lambda_{0})$. Finally, by truncating the infinite series for $D(\lambda
)$\ (\ref{D}) we obtain a polynomial in $\lambda-\lambda_{0}$
\begin{align}
D_{N}(\lambda)  &  =\sum_{n=0}^{N}\left(  \tilde{X}^{(2n)}(T)+X^{(2n)}%
(T)\right)  (\lambda-\lambda_{0})^{n}\label{DN}\\
&  =2+\sum_{n=1}^{N}\left(  \tilde{X}^{(2n)}(T)+X^{(2n)}(T)\right)
(\lambda-\lambda_{0})^{n}.\nonumber
\end{align}
The roots of the polynomials $D_{N}(\lambda)\pm2$ give us eigenvalues
corresponding to Eq.~(\ref{eq}) with periodic and antiperiodic boundary conditions.

As an example, we consider the Mathieu equation with the following
coefficients
$$
p(x)=1,\quad q(x)=2r\cos2x~.
$$
The algorithm was implemented in Matlab 2006. The recursive integration
required for the construction of $\tilde{X}_{0}^{(n)}$, $X_{0}^{(n)}$,
$\tilde{X}^{(n)}$ and $X^{(n)}$ was done by representing the integrand through
a cubic spline using the $\emph{spapi}$ routine with a division of the
interval $[0,T]$ into $7000$ subintervals and integrating using the
$\emph{fnint}$ routine. Next, the zeros of $D_{N}(\lambda)\pm2$\ were
calculated by means of the $\emph{fnzeros}$ routine.

In Tables 4.1 and 4.2, the Mathieu eigenvalues were calculated employing the
SPPS \ representation (\ref{D}) for two values of the parameter $r$. For
comparison the same eigenvalues from the National Bureau of Standards (NBS)
tables are also displayed \cite{NBS}.


\begin{center}
\begin{tabular}
[b]{|c|c|c|}\hline
\multicolumn{3}{|c|}{Table 4.1: $\lambda_{n}$ for the Mathieu Hamiltonian}%
\\\hline
& $r=1$ & $r=1$\\\hline
$n$ & $\lambda_{n}\ \text{(SPPS\thinspace)}$ & $\lambda_{n}%
\ \text{(NBS\thinspace)}\,$\\\hline
$0$ & $-0.455139055973837$ & $-0.45513860$\\\hline
$1$ & $-0.110248420387377$ & $-0.11024882$\\\hline
$2$ & $1.859107160521687$ & $1.85910807$\\\hline
$3$ & $3.917024962694820$ & $3.91702477$\\\hline
$4$ & $4.371299312651704$ & $4.37130098$\\\hline
$5$ & $9.047736927007582$ & $9.04773926$\\\hline
$6$ & $9.078369587941564$ & $9.07836885$\\\hline
$7$ & $16.033018848985410$ & $16.03297008$\\\hline
$8$ & $16.033785039658117$ & $16.03383234$\\\hline
$9$ & $25.020598536509114$ & $25.02084082$\\\hline
$10$ & $25.021087773318282$ & $25.02085434$\\\hline
\end{tabular}
\qquad
\end{center}

\begin{center}%
\begin{tabular}
[b]{|c|c|c|}\hline
\multicolumn{3}{|c|}{Table 4.2: $\lambda_{n}$ for the Mathieu Hamiltonian}%
\\\hline
& $r=5$ & $r=5$\\\hline
$n$ & $\lambda_{n}\ \text{(SPPS\thinspace)}$ & $\lambda_{n}
\ \text{(NBS\thinspace)}\,$\\\hline
$0$ & $-5.800045777242780$ & $-5.80004602$\\\hline
$1$ & $-5.790080596840196$ & $-5.79008060$\\\hline
$2$ & $1.858191484309548$ & $1.85818754$\\\hline
$3$ & $2.099460384254221$ & $2.09946045$\\\hline
$4$ & $7.449142541577460$ & $7.44910974$\\\hline
$5$ & $9.236327731534002$ & \\\hline
$6$ & $11.548906947651728$ & \\\hline
$7$ & $16.648219815375526$ & \\\hline
$8$ & $17.096668282587867$ & \\\hline
$9$ & $25.510753265631860$ & $25.51081605$\\\hline
$10$ & $25.551677357240167$ & $25.54997175$\\\hline
\end{tabular}
\end{center}


Figures 1 and 2 display the plots of the calculated Hill discriminants for two
values of the Mathieu parameter. A supersymmetric Mathieu potential can be
written in terms of the even Mathieu cosine function as follows \cite{Liu01}:
\begin{equation}
\label{susypMathieu}V_{2}=2\left(  \frac{d_{x}Ce(\lambda_{0}%
,r,x)}{Ce(\lambda_{0},r,x)}\right)  ^{2}+2 \lambda_{0}-2r\cos(2x)
\end{equation}
and has the same Hill discriminants for identical values of the parameter $r$.

\begin{center}
\begin{figure}[ptb]
\centering
\includegraphics[width= 16.5 cm, height=6.5 cm]{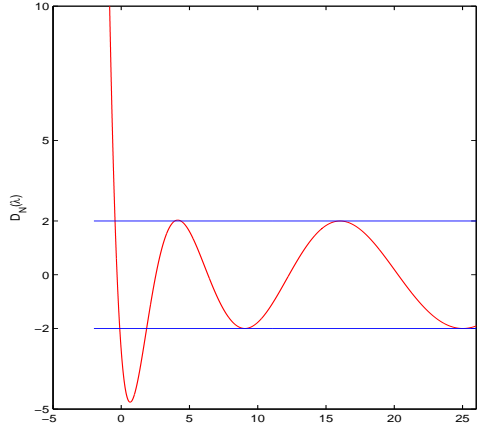}
\caption{\textsl{The polynomial $D_{N}(\lambda)$ for the Mathieu equation with
the parameter $r=1$ calculated by means of formula (\ref{DN}) for $N=100$.}}%
\label{mathi1}%
\end{figure}
\end{center}

\begin{figure}[ptb]
\caption{\textsl{Same as in the previous figure but for $r=5$. The first
minimum goes down to -292.0066.}}%
\label{mathi2}%
\centering
\includegraphics[width= 16.5 cm, height=6.5 cm]{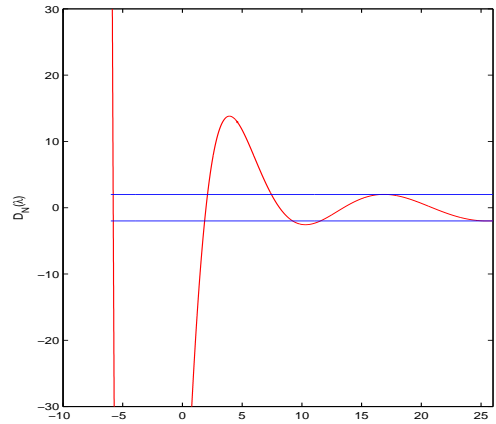}
\end{figure}

As another example, consider the potential
\begin{equation}
\label{V1}V_{1} =\frac{\xi^{2}}{8}\left(  1-\cos4x\right)  -3\xi\cos2x~,
\end{equation}
which belongs to the quasi-exactly solvable family of the so-called
trigonometric Razavy potentials \cite{raz}. The parameter $\xi$ is a positive
real number. In Tables 4.3 and 4.4, the Razavy eigenvalues were calculated
employing the SPPS \ representation (\ref{D}) for two different values of the
parameter $\xi$. For comparison we use the eigenvalues given by Razavy
analytically in terms of the parameter $\xi$ as follows \cite{raz}%
\[
\lambda_{0}=2\left(  1-\sqrt{1+\xi^{2}}\right)  ,\quad\lambda_{3}%
=4,\quad\lambda_{4}=2\left(  1+\sqrt{1+\xi^{2}}\right)  ~.
\]

$
\begin{tabular}
[b]{|c|c|c|}\hline
\multicolumn{3}{|c|}{Table 4.3: $\lambda_{n}$ for the Razavy Hamiltonian}%
\\\hline
& \qquad\quad$\xi=1$ & \qquad\quad$\xi=1$\\\hline
$n$ & \qquad$\lambda_{n}\ \text{(SPPS\thinspace)}$ & \qquad$\lambda
_{n}\ \text{(Ref. \cite{raz}\thinspace)}$\\\hline
$0$ & $-0.828430172936322$ & $-0.828427124746190$\\\hline
$1$ & $-0.627099642286704$ & \\\hline
$2$ & $2.315154289053194$ & \\\hline
$3$ & $3.999948252396118$ & $4$\\\hline
$4$ & $4.834668005757639$ & $4.828427124746190$\\\hline
$5$ & $9.246360795065604$ & \\\hline
$6$ & $9.305957668676312$ & \\\hline
\end{tabular}
\ \qquad$

$%
\begin{tabular}
[b]{|c|c|c|}\hline
\multicolumn{3}{|c|}{Table 4.4: $\lambda_{n}$ for the Razavy Hamiltonian}%
\\\hline
& \qquad\quad$\xi=2$ & \qquad\quad$\xi=2$\\\hline
$n$ & \qquad$\lambda_{n}\ \text{(SPPS\thinspace)}$ & \qquad$\lambda
_{n}\ \text{(Ref. \cite{raz})}\,$\\\hline
$0$ & $-2.472136690058546$ & $-2.472135954999580$\\\hline
$1$ & $-2.428288532265432$ & \\\hline
$2$ & $3.193559545313260$ & \\\hline
$3$ & $4.000042398350143$ & $4$\\\hline
$4$ & $6.472170127477180$ & $6.472135954999580$\\\hline
$5$ & $9.864070609921770$ & \\\hline
$6$ & $10.253303565368553$ & \\\hline
\end{tabular}
\ $

\medskip

\noindent In Figs. 3 and 4 we display the plots of the Hill discriminants for
the values of the Razavy parameter $\xi=1$ and $\xi=2$, respectively.
According to our results in subsection (\ref{ss4-6}), the Hill discriminant is
the same in the case of the supersymmetric partner potential
\begin{equation}
\label{V2-Raz}V_{2} =V_{1} +4\cos2x\left(  \frac{\xi}{2}-\frac{2A(\xi)}%
{\xi-A(\xi) \cos2x}\right)  +\frac{8A(\xi) \sin^{2}2x}{\left(  \xi-A(\xi)
\cos2x\right)  ^{2}}%
\end{equation}
for the same values of the parameter $\xi$. In the latter equation
$A(\xi)=\left(  1-\sqrt{1+\xi^{2}}\right)  $.

\begin{figure}[h]
\par
\begin{center}
\includegraphics[width= 27.5 cm, height=8.5 cm]{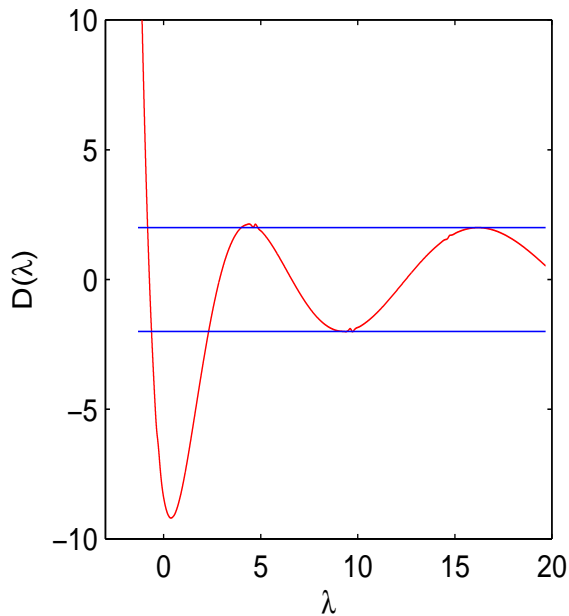}
\end{center}
\caption{\textsl{The polynomial $D_{N}(\lambda)$ for the Razavy equation with
the parameter $\xi=1$ calculated by means of formula (\ref{DN}) for $N=100$.}}%
\label{razF2}%
\end{figure}

\begin{figure}[h]
\par
\begin{center}
\includegraphics[width= 24.5 cm, height=8.5 cm]{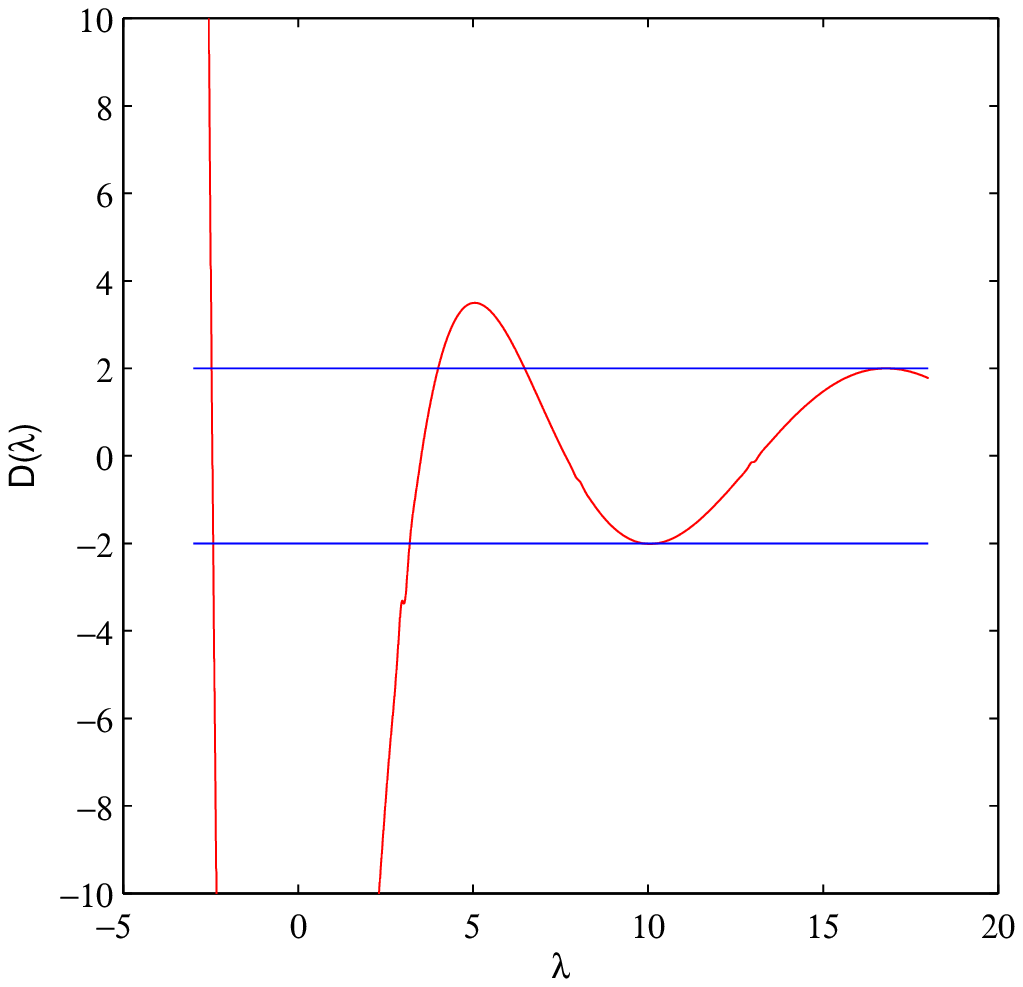}
\end{center}
\caption{\textsl{Same as in the previous figure but for $\xi=2$. The first
minimum of Hill's discriminant goes down to -55.01.}}%
\label{razF3}%
\end{figure}

As final comments to this subsection, we believe that the calculation of the
Hill discriminant through (\ref{DN}) offers clear numerical advantages with
respect to other more complicated formulas for this important quantity
provided in the literature, such as Jagerman's so-called cardinal series
representation \cite{Jagerman}, the infinite determinant representation
involving the Fourier coefficients of the potential as well as the spectral
parameter in the book of Magnus and Winkler \cite{Magnus},
a matrix representation whose entries are complicated phase integrals obtained
by the phase-integral method used by Fr\"{o}man \cite{Froman}, and Boumenir's
representation in terms of integrals derived from the inverse spectral theory
\cite{Boum}.

\section{Spectral and transmission problems on the whole
line\label{SectSpectral and Transmission}}

In this section we consider the one-dimensional Schr\"{o}dinger equation
\begin{equation}
Hu(x)=-u^{\prime\prime}(x)+Q(x)u(x)=\lambda u(x),\quad x\in\mathbb{R},
\label{Hu}%
\end{equation}
where
\begin{equation}
Q(x)=\left\{
\begin{array}
[c]{cc}%
\alpha_{1}, & x<0,\\
q(x), & 0\leq x\leq h,\\
\alpha_{2}, & x>h,
\end{array}
\right.  \label{Q}%
\end{equation}
$\alpha_{1}$ and $\alpha_{2}$ are complex constants and $q$ is a continuous
complex-valued function defined on the segment $[0,h]$. Thus, outside a finite
segment the potential $Q$ admits constant values, and at the end points of the
segment the potential may have discontinuities. We are interested in two
classical problems. The first is the quantum-mechanical spectral problem, we
are looking for such values of the spectral parameter $\lambda\in\mathbb{C}$
for which the Schr\"{o}dinger equation possesses a solution $u$ belonging to
the Sobolev space $H^{2}(\mathbb{R)}$ which in the case of the potential of
the form (\ref{Q}) means that we are looking for solutions exponentially
decreasing at $\pm\infty$.

The second consists in finding the reflectance and transmittance of the
inhomogeneous layer described by $q$. We will formulate this problem in the
form in which it arises in electromagnetic theory though both problems come
not from one but from many different branches of physics and engineering.

\subsection{Quantum-mechanical spectral problem}

The eigenvalue problem (\ref{Hu}) is one of the central in quantum mechanics
for which $H$ is a self-adjoint operator in $L^{2}(\mathbb{R})$ with the
domain $H^{2}(\mathbb{R)}.$ It implies that $Q$ is a real-valued function. In
this case the operator $H$ has a continuous spectrum $[\min\left\{  \alpha
_{1},\alpha_{2}\right\}  ,+\infty)$ and a discrete spectrum located on the
set
\begin{equation}
\lbrack\min_{x\in\lbrack0,h]}q(x),\min\left\{  \alpha_{1},\alpha_{2}\right\}
). \label{interval}%
\end{equation}
Computation of energy levels of a quantum well described by the potential $Q$
is a problem of physics of semiconductor nanostructures (see, e.g.,
\cite{Harrison}). Other important models which reduce to the spectral problem
(\ref{Hu}) arise in studying the electromagnetic and acoustic wave propagation
in inhomogeneous waveguides (see for instance \cite{BC}, \cite{Ch}, \cite{FM},
\cite{WCC}, \cite{Breh}, \cite{OR}, \cite{MC}).

Hence in the applied problems it is important to have effective and rapid
numerical methods for the solution of the problem (\ref{Hu}). The most
frequently applied is the shooting method (see, e.g., \cite{Harrison}). It has
well known limitations due to\ the intrinsic difficulties of the shooting
procedure, especially when the spectral parameter as in the problem under
consideration participates in the boundary conditions (see equalities
(\ref{cond0}) and (\ref{disp1}) below). It is much more convenient to have
available an analytical form of a dispersion equation associated with the
eigenvalue problem. In that case solutions of the dispersion equation can be
approximated using different numerical techniques. However the dispersion
equation is available only in really few examples (see \cite{Flugge}). There
is another method developed for symmetric potentials \cite{Hall}. Below we
compare numerical results of our approach with the results from \cite{Hall}.

For simplicity we will assume that $\alpha_{1}$ and $\alpha_{2}$ are real
constants and the function $q$ is a continuous real-valued function on $[0,h]$
though the presented method is applicable to the more general situation when
$Q$ is complex valued. In this case necessary modifications must be made
mainly in the reduction of the original problem on the whole line to a problem
for the equation
\begin{equation}
-u^{\prime\prime}(x)+q(x)u(x)=\lambda u(x),\quad x\in(0,h) \label{eq0h}%
\end{equation}
with three boundary conditions at the end points of the interval $(0,h)$ (see
below) meanwhile the application of the SPPS method suffers no essential
changes. Our analysis follows that from \cite{CKOR}.

For $x<0$ we have to consider the equation $-u^{\prime\prime}+(\alpha
_{1}-\lambda)u=0$. Its solutions decreasing at $-\infty$ exist if only
$\alpha_{1}-\lambda>0$. Denote $\mu=+\sqrt{\alpha_{1}-\lambda}$. Then the
required solution for $x<0$ has the form $u(x)=c_{1}e^{\mu x}$ and the
multiplicative constant can always be chosen equal to one. Thus, $u(x)=e^{\mu
x}$, $x<0$, from where
\begin{equation}
u(0)=1\quad\text{and}\quad u^{\prime}(0)=\mu. \label{cond0}%
\end{equation}
This gives us the initial conditions for the solution on the interval $(0,h)$
which we will construct following Theorem \ref{ThSolSL}. For that we need
first a nonvanishing particular solution of the equation
\begin{equation}
-u_{0}^{\prime\prime}(x)+q(x)u_{0}(x)=0 \label{SchrHom}%
\end{equation}
which as was explained in Section \ref{Sect Sol S-L} can be constructed by
means of the same SPPS method. Indeed, formulas (\ref{v1v2})-(\ref{Y}) where
$p$ should be chosen equal to $-1$ and $x_{0}=0$ give us a couple of linearly
independent real-valued particular solutions $v_{1}$ and $v_{2}$ of
(\ref{SchrHom}). Hence (see Remark \ref{Rem Sturm separation theorem}) the
required nonvanishing solution of (\ref{SchrHom}) can be chosen as
$u_{0}=v_{1}+iv_{2}$. Let us notice that as $u_{0}(0)=1$ and $u_{0}^{\prime
}(0)=i$ the initial conditions satisfied by the solutions of (\ref{eq0h})
$u_{1}$ and $u_{2}$ constructed according to (\ref{gensol}) have the form%
\[
u_{1}(0)=u_{0}(0)=1,\quad u_{1}^{\prime}(0)=u_{0}^{\prime}(0)=i,
\]%
\[
u_{2}(0)=0,\quad u_{2}^{\prime}(0)=-\frac{1}{u_{0}(0)}=-1.
\]
From these relations we obtain that the solution of (\ref{eq0h}) satisfying
the initial conditions (\ref{cond0}) has the form%
\begin{equation}
u(x)=u_{1}(x)+(i-\mu)u_{2}(x)\quad0\leq x\leq h. \label{solinside}%
\end{equation}

In the region $x>h$ the solution of equation (\ref{Hu}) has the form
\[
u(x)=C_{1}e^{-\sqrt{\alpha_{2}-\lambda}(x-h)}+C_{2}e^{\sqrt{\alpha_{2}%
-\lambda}(x-h)}%
\]
from which we obtain that the existence of an eigenfunction is possible if
only $\sqrt{\alpha_{2}-\lambda}\in\mathbb{R}$. Hence $\alpha_{2}>\lambda$ and
we denote $\nu=+\sqrt{\alpha_{2}-\lambda}$. Consequently, $u(x)=Ce^{-\nu
(x-h)}$ and $u(h)=C$, $u^{\prime}(h)=-\nu C$ where $C$ is an arbitrary
constant. Thus, the eigenvalues of the problem are such values of $\lambda$
for which the solution (\ref{solinside}) satisfies the condition
\begin{equation}
u^{\prime}(h)+\nu u(h)=0 \label{disp1}%
\end{equation}
where, as above, $\nu=+\sqrt{\alpha_{2}-\lambda}$ and $\alpha_{2}>\lambda$.

In order to write down the explicit form of the dispersion equation
(\ref{disp1}) in terms of the spectral parameter power series we calculate the
derivatives of the solutions of (\ref{eq0h}),%
\[
u_{1}^{\prime}=\frac{u_{0}^{\prime}}{u_{0}}u_{1}-\frac{\lambda}{u_{0}}%
{\displaystyle\sum\limits_{n=0}^{\infty}}
\lambda^{n}\widetilde{X}^{(2n+1)}\quad\text{and}\quad u_{2}^{\prime}%
=\frac{u_{0}^{\prime}}{u_{0}}u_{2}-\frac{1}{u_{0}}%
{\displaystyle\sum\limits_{n=0}^{\infty}}
\lambda^{n}X^{(2n)}.
\]
Thus the derivative of the solution (\ref{solinside}) has the form%
\[
u^{\prime}=\frac{u_{0}^{\prime}}{u_{0}}u-\frac{1}{u_{0}}\left(
{\displaystyle\sum\limits_{n=0}^{\infty}}
\lambda^{n+1}\widetilde{X}^{(2n+1)}+(i-\mu)%
{\displaystyle\sum\limits_{n=0}^{\infty}}
\lambda^{n}X^{(2n)}\right)  .
\]
Substituting this expression into (\ref{disp1}) we arrive at the following
result obtained in \cite{CKOR} and formulated here in the form of a theorem.

\begin{theorem}
Let $\alpha_{1}$, $\alpha_{2}$ be real numbers, $q$ be a real-valued
continuous function defined on $[0,h]$ and $Q$ be defined by (\ref{Q}). Then
$\lambda\in\lbrack\min_{x\in\lbrack0,h]}q(x),\min\left\{  \alpha_{1}%
,\alpha_{2}\right\}  )$ is an eigenvalue of the problem (\ref{Hu}) if and only
if the following dispersion equation
\begin{gather}
u_{0}^{\prime}(h)\left(
{\displaystyle\sum\limits_{n=0}^{\infty}}
\lambda^{n}\widetilde{X}^{(2n)}(h)+\left(  i-\sqrt{\alpha_{1}-\lambda}\right)
%
{\displaystyle\sum\limits_{n=0}^{\infty}}
\lambda^{n}X^{(2n+1)}(h)\right) \nonumber\\
-\frac{1}{u_{0}(h)}\left(
{\displaystyle\sum\limits_{n=0}^{\infty}}
\lambda^{n+1}\widetilde{X}^{(2n+1)}(h)+\left(  i-\sqrt{\alpha_{1}-\lambda
}\right)
{\displaystyle\sum\limits_{n=0}^{\infty}}
\lambda^{n}X^{(2n)}(h)\right) \nonumber\\
+\sqrt{\alpha_{2}-\lambda}u_{0}(h)\left(
{\displaystyle\sum\limits_{n=0}^{\infty}}
\lambda^{n}\widetilde{X}^{(2n)}(h)+\left(  i-\sqrt{\alpha_{1}-\lambda}\right)
%
{\displaystyle\sum\limits_{n=0}^{\infty}}
\lambda^{n}X^{(2n+1)}(h)\right)  =0~, \label{dispersionGeneral1}%
\end{gather}
is satisfied and the corresponding (unique up to a multiplicative constant)
eigenfunction has the form%
\[
u(x)=\left\{
\begin{array}
[c]{cc}%
e^{\mu x}, & x<0,\\
u_{1}(x)+(i-\mu)u_{2}(x), & 0\leq x\leq h,\\
\left(  u_{1}(h)+(i-\mu)u_{2}(h)\right)  e^{-\nu(x-h)}, & x>h~,
\end{array}
\right.
\]
where $\mu=+\sqrt{\alpha_{1}-\lambda}$, $\nu=+\sqrt{\alpha_{2}-\lambda}$ and
$u_{1}$, $u_{2}$ are defined by (\ref{gensol}) where $u_{0}$ is the
nonvanishing solution of (\ref{SchrHom}) on $(0,h)$ satisfying the initial
conditions $u_{0}(0)=1$ and $u_{0}^{\prime}(0)=i$, $p\equiv-1$, $r\equiv1$ and
$x_{0}=0$.
\end{theorem}

All the coefficients in equation (\ref{dispersionGeneral1}): $u_{0}(h)$,
$u_{0}^{\prime}(h)$, $\widetilde{X}^{(k)}(h)$ and $X^{(k)}(h)$ are easily and
(as our numerical tests show) accurately obtained from the definitions
introduced above, and the roots of the dispersion equation coincide with the
eigenvalues of the problem and can be found using many available methods.

In what follows, let us consider a relatively simple situation: $\alpha
_{1}=\alpha_{2}$. Rearranging the terms in equation (\ref{Hu}) this case
always can be reduced to the case $\alpha_{1}=\alpha_{2}=0$. Then $\nu
=\mu=\sqrt{-\lambda}$,\ $\mu^{2}=-\lambda$ and $\lambda^{n}=(-1)^{n}\mu^{2n}$.
The dispersion equation takes the form (here we correct some easily detectable
misprints in \cite{CKOR})
\begin{gather*}
u_{0}^{\prime}(h)\left(  1+iX^{(1)}(h)\right)  -\frac{i}{u_{0}(h)}\\
+%
{\displaystyle\sum\limits_{n=1}^{\infty}}
(-1)^{n}\mu^{2n}(u_{0}^{\prime}(h)\widetilde{X}^{(2n)}(h)-\frac{1}{u_{0}%
(h)}\widetilde{X}^{(2n-1)}(h)+iu_{0}^{\prime}(h)X^{(2n+1)}(h)\\
-\frac{i}{u_{0}(h)}X^{(2n)}(h)+u_{0}(h)X^{(2n-1)}(h))\\
+%
{\displaystyle\sum\limits_{m=0}^{\infty}}
(-1)^{m}\mu^{2m+1}(-u_{0}^{\prime}(h)X^{(2m+1)}(h)+\frac{1}{u_{0}(h)}%
X^{(2m)}(h)\\
+iu_{0}(h)X^{(2m+1)}(h)+u_{0}(h)\widetilde{X}^{(2m)}(h))=0.
\end{gather*}

Thus, the dispersion equation has the form%
\begin{equation}
\sum_{k=0}^{\infty}a_{k}\mu^{k}=0 \label{disp}%
\end{equation}
where
\begin{equation}
a_{0}=u_{0}^{\prime}(h)\left(  1+iX^{(1)}(h)\right)  -\frac{i}{u_{0}(h)},
\label{a0}%
\end{equation}%
\begin{gather}
a_{2n}=(-1)^{n}(u_{0}^{\prime}(h)\widetilde{X}^{(2n)}(h)-\frac{1}{u_{0}%
(h)}\widetilde{X}^{(2n-1)}(h)+iu_{0}^{\prime}(h)X^{(2n+1)}(h)\nonumber\\
-\frac{i}{u_{0}(h)}X^{(2n)}(h)+u_{0}(h)X^{(2n-1)}(h)),\quad n\in\mathbb{N},
\label{a2n}%
\end{gather}

\begin{gather}
a_{2n+1}=(-1)^{n}(-u_{0}^{\prime}(h)X^{(2m+1)}(h)+\frac{1}{u_{0}(h)}%
X^{(2m)}(h)\nonumber\\
+iu_{0}(h)X^{(2m+1)}(h)+u_{0}(h)\widetilde{X}^{(2m)}(h)),\quad n=0,1,2,\ldots.
\label{a2n+1}%
\end{gather}
The problem is reduced to the problem of finding zeros of an analytic function
given by its Taylor series with the coefficients $a_{k}$, $k=0,1,2,\ldots$.

The usual approach to numerical solution of the considered eigenvalue problem
consists in applying the shooting method (see, e.g., \cite{Harrison}) which is
known to be unstable, relatively slow and to the difference of our approach
does not offer any explicit equation for determining eigenvalues and
eigenfunctions. In \cite{Hall} another method based on approximation of the
potential by square wells was proposed. It is limited to the case of symmetric
potentials. The approach based on the SPPS is completely different and does
not require any shooting procedure, approximation of the potential or
numerical differentiation. Derived from the exact dispersion equation
(\ref{disp}) we consider its approximation $\sum_{k=0}^{N}a_{k}\mu^{k}=0$ and
in fact look for zeros of the polynomial $\sum_{k=0}^{N}a_{k}\mu^{k}$ in the
interval $[\min q(x),0)$. Here we give only one example of numerical
computation of eigenvalues referring to \cite{CKOR} for more examples and discussion.

We consider the potential $Q$ defined by the expression $Q(x)=-\upsilon
\operatorname{sech}^{2}x$, $x\in(-\infty,\infty)$. It is not of a finite
support, nevertheless its absolute value decreases rapidly when $x\rightarrow
\pm\infty$. We approximate the original problem by a problem with a finite
support potential $\widehat{Q}$ defined by the equality%
\[
\widehat{Q}(x)=\left\{
\begin{array}
[c]{cc}%
0, & x<-a\\
-\upsilon\operatorname{sech}^{2}x, & -a\leq x\leq a\\
0, & x>a~.
\end{array}
\right.
\]

An attractive feature of the potential $Q$ is that its eigenvalues can be
calculated explicitly (see, e.g., \cite{Flugge}). In particular, for
$\upsilon=m(m+1)$ the eigenvalue $\lambda_{n}$ is given by the formula
$\lambda_{n}=-(m-n)^{2}$, $n=0,1,\ldots$. \bigskip

The results of application of the SPPS method for $\upsilon=12$ are given in
Table \ref{Table Sech} in comparison with the exact values and the results
from \cite{Hall}.

\begin{center}

\label{Table Sech}%

\begin{tabular}
[c]{|c|c|c|c|}\hline
\multicolumn{4}{|c|}{Table 5.1: Approximations of $\lambda_{n}$ of the
Hamiltonian $H=-D^{2}-12\operatorname{sech}^{2}x$ \ }\\\hline
$n$ & Exact values & Numerical results from \cite{Hall} & Numerical results
using SPPS ($N=180$)\\\hline
0 & $-9$ & $-9.094$ & $-8.999628656$\\\hline
1 & $-4$ & $-4.295$ & $-3.999998053$\\\hline
2 & $-1$ & $-0.885$ & $-0.999927816$\\\hline
\end{tabular}

\bigskip
\end{center}

\subsection{Transmission problem for inhomogeneous layers}

In this subsection we apply the SPPS method to the problem of finding the
reflectance and transmittance of a finite inhomogeneous layer. This is a
classical problem which still attracts a lot of attention due to its numerous
applications in modern engineering, optical physics, solution of nonlinear
problems and many other fields. Different methods for numerical solution of
the problem have been proposed, mainly based on well known canonical
techniques for approximate solution of ordinary differential equations such as
the finite differences or expansion in power series (see, e.g.,
\cite{Kildemo1997}, \cite{Yeh}, \cite{Khalaj}, \cite{Chamanzar2006}). One of
the most used methods involves the approximation of the inhomogeneous layer by
a structure consisting of many homogeneous layers (see, e.g., \cite{Knittl},
\cite{Montecchi95}, \cite{Caro}, \cite{Su2008}). Asymptotic methods such as
the perturbation method or the WKB method are also applied to this problem
(see, e.g., \cite{Hall1958}, \cite{Tikhonravov et al}, \cite{Montecchi2001},
\cite{Yeh}), though in the case of a finite inhomogeneous layer the WKB
technique does not seem advantageous. Meanwhile the mentioned numerical
approaches can give satisfactory results for certain fixed parameters of the
problem their applicability is questionable when the solution of the problem
is required, for example, for many different angles of incidence. The
treatment of the oblique incidence case is not only interesting because of the
many applications in which that incidence is needed - in optical filters,
light couplers -, but also because sometimes the interfaces are rough - their
effects and analysis depending on their size -, and/or are not parallel (see,
e.g., \cite{Montecchi2001}). This is due to imperfect deposition conditions.
Such problems in the generation of the inhomogeneous layer (or multilayer)
have generated systems in which the feedback of a reflectance, transmittance,
or scattered light measurement is used to characterize the layer as it is
created and to correct any discrepancies with the pre-established values. Such
application must be able to recalculate the required correction profile and
requires a real-time computation of transmittance and reflectance.

\begin{figure}
[ptb]
\begin{center}
\includegraphics[
height=2.4388in,
width=3.7827in
]%
{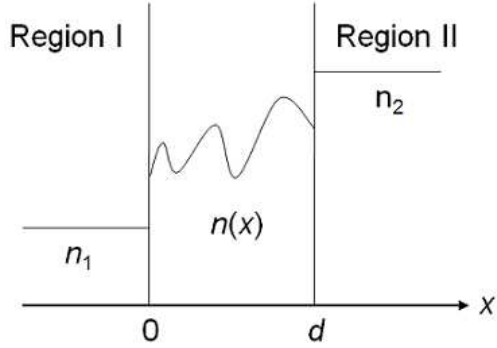}
\caption{An inhomogeneous layer.}%
\label{Fig1}%
\end{center}
\end{figure}

The mathematical statement of the problem involves a Helmholtz equation with a
coefficient which is an arbitrary continuous function on a finite segment and
constant outside. More precisely, the scalar function $u$ which represents a
component of a linearly polarized electromagnetic wave in the case of an
$s$-polarization satisfies the Helmholtz equation
\begin{equation}
u^{^{\prime\prime}}(x)+[k^{2}n^{2}(x)-\beta^{2}]u(x)=0 \label{HelmholtzBeta}%
\end{equation}
where $u$ stands for the transverse component of the electric field, and in
the case of a $p$-polarization satisfies the following Sturm-Liouville
equation (see, e.g., \cite{Ishimaru}, \cite{Felbacq})%
\begin{equation}
n^{2}(x)\left(  \frac{1}{n^{2}(x)}v^{\prime}(x)\right)  ^{\prime}+[k^{2}%
n^{2}(x)-\beta^{2}]v(x)=0 \label{SLBeta}%
\end{equation}
in which $v$ represents the transverse component of the magnetic field. Here
$k$ is the free-space circular wave number. The refractive index $n$ preserves
constant values $n_{1}$ and $n_{2}$ in the regions $x<0$ and $x>d$
respectively and is an arbitrary continuous function in the interval $0\leq
x\leq d$ (see figure \ref{Fig1}).\ For simplicity we assume $n$ to be real
valued though the method is equally applicable to the case of a complex
refractive index.

The propagation constant $\beta$ is related to the angle of incidence of the
wave in the following way $\beta=k\sin\theta$ (see, e.g., \cite{Ishimaru}),
and $\beta$ vanishes in the case of normal incidence.%

In spite of the fact that equations (\ref{HelmholtzBeta}) and (\ref{SLBeta})
describe the behaviour of different components of an electromagnetic wave,
corresponding to an electric and a magnetic field respectively, there exists a
simple transformation from (\ref{SLBeta}) to (\ref{HelmholtzBeta}) and vice
versa (see, e.g., \cite{Ishimaru}). Namely, if $v$ is a solution of
(\ref{SLBeta}) then $U=v/n$ is a solution of the equation
\[
U^{^{\prime\prime}}(x)+[k^{2}N^{2}(x)-\beta^{2}]U(x)=0
\]
where $k^{2}N^{2}=k^{2}n^{2}+n^{^{\prime\prime}}/n-2\left(  n^{\prime
}/n\right)  ^{2}$. Thus, in both cases the problem reduces to an equation of
the form (\ref{HelmholtzBeta}).

\begin{figure}
[ptb]
\begin{center}
\includegraphics[
trim=0.000000in 0.000000in -0.000591in 0.000000in,
height=2.5789in,
width=3.9505in
]%
{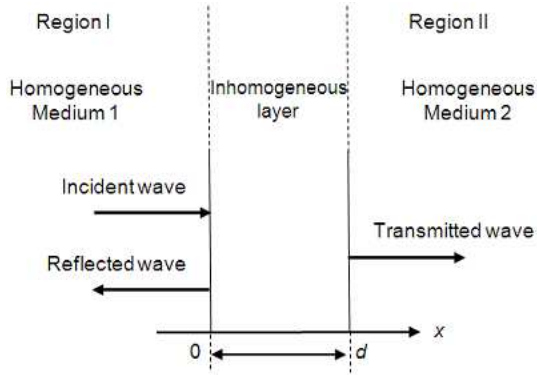}
\caption{Incident, reflected and transmitted waves.}%
\label{Fig2}%
\end{center}
\end{figure}

We denote $k_{1}=\sqrt{k^{2}n_{1}^{2}-\beta^{2}}$ and $k_{2}=\sqrt{k^{2}%
n_{2}^{2}-\beta^{2}}$. The solution $u$ of (\ref{HelmholtzBeta}) or $v$ of
(\ref{SLBeta}) respectively together with their first derivatives must be
continuous at all $x$ including the points $x=0$ and $x=d$. The incident wave
in the region I (see figure \ref{Fig2})
is assumed to have the form $e^{-ik_{1}x},$ and together with the reflected
wave the whole solution for $x<0$ is the combination
\[
u(x)=e^{-ik_{1}x}+Re^{ik_{1}x},\quad x<0
\]
where the constant $R$ is the reflection coefficient whose absolute value is
less than 1. The solution corresponding to the transmitted wave in the region
II has the form
\[
u(x)=Te^{-ik_{2}x},\quad x>d
\]
where $T$ is the transmission coefficient. In the case of unabsorbent media
for the normally incident waves the following energy conservation relation
holds
\begin{equation}
\left\vert R\right\vert ^{2}+n_{2}\left\vert T\right\vert ^{2}/n_{1}=1.
\label{EnergyConserved}%
\end{equation}

Let us suppose that the two linearly independent solutions $y_{1}$ and $y_{2}$
of (\ref{HelmholtzBeta}) in the interval of inhomogenicity $0\leq x\leq d$ are
known such that the following initial conditions are satisfied:
\begin{equation}
y_{1}(0)=1,\quad y_{1}^{\prime}(0)=0 \label{u(0) uno}%
\end{equation}
and
\begin{equation}
y_{2}(0)=0,\quad y_{2}^{\prime}(0)=1. \label{u(0) dos}%
\end{equation}
Then we are able to obtain analytic expressions for $R$ and $T$ \ in terms of
$u_{1}$ and $u_{2}$ \cite{CKKO2009}. We have
\begin{equation}
R=\frac{-k_{1}k_{2}y_{2}(d)-y_{1}^{\prime}(d)-ik_{2}y_{1}(d)+ik_{1}%
y_{2}^{\prime}(d)}{[y_{1}^{\prime}(d)-k_{1}k_{2}y_{2}(d)]+i[k_{2}%
y_{1}(d)+k_{1}y_{2}^{\prime}(d\text{ })]} \label{R}%
\end{equation}
and
\begin{equation}
T=\frac{2ik_{1}[y_{1}(d)y_{2}^{\prime}(d)-y_{1}^{\prime}(d)y_{2}%
(d)]e^{ik_{2}d}}{[y_{1}^{\prime}(d)-k_{1}k_{2}y_{2}(d)]+i[k_{2}y_{1}%
(d)+k_{1}y_{2}^{\prime}(d)]}. \label{T}%
\end{equation}
These formulas remain valid for equation (\ref{SLBeta}) when one substitutes
$y_{1}$ and $y_{2}$ with the solutions $v_{1}$ and $v_{2}$ of (\ref{SLBeta})
satisfying the initial conditions (\ref{u(0) uno}) and (\ref{u(0) dos}) respectively).

Thus, the transmission problem for an inhomogeneous layer consists in
computing a couple of solutions of (\ref{HelmholtzBeta}) (or (\ref{SLBeta}))
in the interval of inhomogenicity $0\leq x\leq d$, satisfying the initial
conditions (\ref{u(0) uno}) and (\ref{u(0) dos}), and then the reflection and
transmission coefficients are found from (\ref{R}) and (\ref{T}). For
computation of these solutions we use theorem \ref{ThSolSL} and take into
account (\ref{at01}) and (\ref{at02}) where it is convenient to choose
$x_{0}=0$.

There are several examples of explicitly solvable inhomogeneous profiles
\cite{Monaco}, \cite{Yeh}. These were used in \cite{CKKO2009} for testing the
results obtained by means of SPPS. In all numerical simulations the achieved
accuracy was remarkable.

\section{Zakharov-Shabat eigenvalue problem\label{Sect Z-S}}

In this section we study the Zakharov-Shabat system with a real-valued
potential. It arises in the solution via the inverse scattering method of
several nonlinear evolution equations such as the nonlinear Schr\"{o}dinger
equation, the sine-Gordon equation and the modified Korteweg-de Vries
equation. For example, in the case of the nonlinear Schr\"{o}dinger equation,
eigenvalues of the Zakharov-Shabat system correspond to soliton solutions
implemented in fiber optics (see, e.g., \cite{Hasegawa}). The assumption that
the potential is real valued is natural and common in the engineering
literature,- it includes the conventional profiles such as the rectangular,
the Gaussian and the hyperbolic secant.

In \cite{KV2011} a general solution of the Zakharov-Shabat system with a real
potential in terms of SPPS was obtained and used for deriving a dispersion
equation corresponding to the eigenvalue problem with a compactly supported
potential. Once again the problem is reduced to a problem of localizing zeros
of an analytic function given by its Taylor series. For numerical
approximation of eigenvalues one can consider a truncated series and thus for
practical computation the eigenvalue problem reduces to finding roots of a polynomial.

The Zakharov-Shabat system with a real potential has the form \cite{ZS1},
\cite{ZS}
\begin{align}
\partial n_{1}(x)-\lambda n_{1}(x)  &  =U(x)n_{2}(x)\label{ZS1}\\
\partial n_{2}(x)+\lambda n_{2}(x)  &  =-U(x)n_{1}(x)~, \label{ZS2}%
\end{align}
where $\partial:=\frac{d}{dx}$; $U:\mathbb{R}\rightarrow\mathbb{R}$ is the
potential and $U\in L_{1}(-\infty,\infty)$; the solutions $n_{1}$ and $n_{2}$
in general are complex valued and the spectral parameter $\lambda$ is a
complex constant. It is convenient to rewrite the Zakharov-Shabat system using
the following notations%

\[
u=n_{1}+in_{2},\qquad v=n_{1}-in_{2},\qquad q=iU~.
\]
Then (\ref{ZS1}), (\ref{ZS2}) takes the form of a Dirac system with a scalar
potential (see, e.g., \cite{Casahorran}, \cite{Hiller2002}, \cite{Ho},
\cite{NogamiToyama1993})
\begin{equation}
(\partial+q(x))u=\lambda v, \label{D1}%
\end{equation}%
\begin{equation}
(\partial-q(x))v=\lambda u~. \label{D2}%
\end{equation}

From these equalities it is easy to see that $u$ and $v$ are solutions of the
following second-order differential equations%
\begin{equation}
(\partial-q(x))(\partial+q(x))u(x)=\lambda^{2}u(x) \label{SchrU}%
\end{equation}
and%

\begin{equation}
(\partial+q(x))(\partial-q(x))v(x)=\lambda^{2}v(x)~. \label{SchrV}%
\end{equation}
The differential operators on the left-hand side can be written in the form of
stationary Schr\"{o}dinger operators describing supersymmetric partners
\[
(\partial-q)(\partial+q)=\partial^{2}+(\partial q-q^{2})\quad\text{and}%
\quad(\partial+q)(\partial-q)=\partial^{2}-(\partial q+q^{2}).
\]
Nevertheless, precisely the factorized form (\ref{SchrU}), (\ref{SchrV})
presents certain advantage for applying the SPPS method due to the possibility
to write down closed-form solutions of (\ref{SchrU}) and (\ref{SchrV}) for
$\lambda=0$. Namely, let $Q(x)=\int q(x)dx$. Then $u_{0}(x)=e^{-Q(x)}$ and
$v_{0}(x)=e^{Q(x)}$ are solutions of the equations $(\partial-q)(\partial
+q)u_{0}=0$ and $(\partial+q)(\partial-q)v_{0}=0$ respectively. Note that for
a continuous function $q$ defined on a closed finite interval both $u_{0}$ and
$v_{0}$ are devoid of zeros.

The systems of auxiliary functions $\left\{  X^{(n)}\right\}  _{n=0}^{\infty}$
and $\left\{  \widetilde{X}^{(n)}\right\}  _{n=0}^{\infty}$ in this case are
defined as follows%
\begin{equation}
X^{(0)}(x)\equiv\tilde{X}^{(0)}(x)\equiv1, \label{X0 ZS}%
\end{equation}%
\begin{equation}
X^{(n)}(x)=\int_{x_{0}}^{x}X^{(n-1)}(s)e^{(-1)^{n}2Q(s)}ds, \label{Xn ZS}%
\end{equation}%
\begin{equation}
\tilde{X}^{(n)}(x)=\int\limits_{x_{0}}^{x}\tilde{X}^{(n-1)}(s)e^{(-1)^{n+1}%
2Q(s)}ds. \label{Xtiln ZS}%
\end{equation}
We obtain the following SPPS form of a general solution of the Zakharov-Shabat system.

\begin{theorem}
\label{teo2}\cite{KV2011} Let $U$ be a continuous real-valued function defined
on a finite segment $[a,b]\subset\mathbb{R}$. Then the general solution of the
Zakharov-Shabat system (\ref{ZS1}), (\ref{ZS2}) has the form%
\begin{equation}
n_{1}(x)=\dfrac{c_{1}}{2}\sum_{n=0}^{\infty}e^{(-1)^{n}Q(x)}\lambda^{n}%
\tilde{X}^{(n)}(x)+\dfrac{c_{2}}{2\lambda}\sum_{n=0}^{\infty}e^{(-1)^{n+1}%
Q(x)}\lambda^{n}X^{(n)}(x), \label{n1}%
\end{equation}%
\begin{equation}
n_{2}(x)=\dfrac{ic_{1}}{2}\sum_{n=0}^{\infty}(-1)^{n}e^{(-1)^{n}Q(x)}%
\lambda^{n}\tilde{X}^{(n)}(x)-\dfrac{ic_{2}}{2\lambda}\sum_{n=0}^{\infty
}(-1)^{n}e^{(-1)^{n+1}Q(x)}\lambda^{n}X^{(n)}(x), \label{n2}%
\end{equation}
where $c_{1}$ and $c_{2}$ are arbitrary complex constants, $Q$ is an
antiderivative of $q=iU$, $x_{0}\in\lbrack a,b]$ and the series converge
uniformly in $[a,b]$.
\end{theorem}

Solutions of the Zakharov-Shabat system (\ref{ZS1}), (\ref{ZS2}) satisfying
the following asymptotic relations
\begin{align}
\vec{\sigma}(x,\lambda)  &  \cong\binom{1}{0}e^{\lambda x}\text{,\quad
\ }x\rightarrow-\infty,\label{jostmeninf}\\
\vec{\xi}(x,\lambda)  &  \cong\binom{0}{1}e^{-\lambda x}\text{,\quad
}x\rightarrow\infty\label{jostinf}%
\end{align}
are called the Jost solutions \cite{Jost}. The eigenvalue problem for the
Zakharov-Shabat system with a real valued potential consists in finding such
values of the spectral parameter $\lambda$ for which $\operatorname*{Re}%
\lambda>0$ and there exists a nontrivial solution $\vec{n}$ satisfying the
Jost conditions (\ref{jostmeninf}) and (\ref{jostinf}).

If the real valued potential $U$ has a compact support on the segment $[-a,a]$
it is easy to see that the eigenvalue problem reduces to find such values of
$\lambda$ ($\operatorname*{Re}\lambda>0$) for which there exists a solution of
(\ref{ZS1}), (\ref{ZS2}) satisfying the following boundary conditions%
\begin{align}
n_{1}(-a)=1,  &  \hspace{1cm}n_{2}(-a)=0,\label{jost+a}\\
n_{1}(a)=0.  &  \label{jost-a}%
\end{align}

We refer here to \cite{Jorobas} and \cite{Potencial} for estimates of the
number of real eigenvalues of a compactly supported potential.

The next statement gives us a dispersion equation equivalent to the
Zakharov-Shabat eigenvalue problem for the real, compactly supported potentials.

\begin{theorem}
\label{eigenvalores}\cite{KV2011} Let $U$ be a continuous real-valued function
with a compact support on the segment $[-a,a]$. Then $\lambda$
($\operatorname*{Re}\lambda>0$) is an eigenvalue of the Zakharov-Shabat system
if and only if the following equation is satisfied%
\begin{equation}
\sum\limits_{n=0}^{\infty}\lambda^{n}\left(  e^{(-1)^{n}Q(a)}\tilde{X}%
^{(n)}(a)+e^{(-1)^{n+1}Q(a)}X^{(n)}(a)\right)  =0~, \label{cte}%
\end{equation}
where $Q(x)=i\int\limits_{-a}^{x}U(t)dt$ and $x_{0}=-a$ in (\ref{X0}%
)-(\ref{Xtil}).

If $\lambda$ is an eigenvalue then the corresponding eigenvector is given by
\[
\overrightarrow{n}=\vec{\psi}+\vec{\varphi}%
\]
with
\begin{align}
\vec{\psi}(x)=\binom{\psi_{1}(x)}{\psi_{2}(x)}=  &  \left(
\begin{array}
[c]{c}%
\dfrac{1}{2}\sum\limits_{n=0}^{\infty}e^{(-1)^{n}Q(x)}\lambda^{n}\tilde
{X}^{(n)}(x)\\
\dfrac{i}{2}\sum\limits_{n=0}^{\infty}(-1)^{n}e^{(-1)^{n}Q(x)}\lambda
^{n}\tilde{X}^{(n)}(x)
\end{array}
\right)  ,\label{Sol n1}\\
\vec{\varphi}(x)=\binom{\varphi_{1}(x)}{\varphi_{2}(x)}=  &  \left(
\begin{array}
[c]{c}%
\dfrac{1}{2}\sum\limits_{n=0}^{\infty}e^{(-1)^{n+1}Q(x)}\lambda^{n}%
X^{(n)}(x)\\
-\dfrac{i}{2}\sum\limits_{n=0}^{\infty}(-1)^{n}e^{(-1)^{n+1}Q(x)}\lambda
^{n}X^{(n)}(x)
\end{array}
\right)  . \label{Sol n2}%
\end{align}

\end{theorem}

The theorem reduces the Zakharov-Shabat eigenvalue problem with a compactly
supported potential to the problem of localizing zeros (in the right
half-plane) of an analytic function $\kappa(\lambda)=\sum\limits_{n=0}%
^{\infty}a_{n}\lambda^{n}$ of the complex variable $\lambda$ with the Taylor
coefficients $a_{n}$ given by the expressions
\begin{equation}
a_{n}=e^{(-1)^{n}Q(a)}\tilde{X}^{(n)}(a)+e^{(-1)^{n+1}Q(a)}X^{(n)}(a).
\label{an}%
\end{equation}
Equation~(\ref{cte}) represents a dispersion equation of the eigenvalue
problem. The coefficients $a_{n}$ can be easily and accurately calculated
following the definitions introduced above. For the numerical solution of the
eigenvalue problem one can consider a polynomial
\begin{equation}
\kappa_{N}(\lambda)=\sum\limits_{n=0}^{N}a_{n}\lambda^{n} \label{kN}%
\end{equation}
approximating the function $\kappa$. For a reasonably large $N$ its roots give
an accurate approximation of the eigenvalues of the problem.

As a numerical example we consider the rectangular box referring the reader to
\cite{KV2011} for further numerical tests and details. For the rectangular box
the exact solution satisfying the boundary conditions (\ref{jost+a}) \ is
known. Such system can be applied to describe the problem of the diffraction
of a wave by a screen with a slit (see \cite{RectaProfManakov}). Discrete
eigenvalues of the spectral parameter can also be approximated using a
variational principle approach \cite{RectaProfDJK}. The potential is defined
by the equality%
\[
U(x)=\left\{
\begin{array}
[c]{ll}%
A, & \left\vert x\right\vert <a,\\
0, & \text{elsewhere.}%
\end{array}
\right.
\]
A dispersion equation in this case can be obtained explicitly and written as
follows%
\begin{equation}
\gamma\cos2\gamma+\lambda\sin2\gamma=0~, \label{trascendental}%
\end{equation}
where $\gamma=\sqrt{A^{2}-\lambda^{2}}$.

For solving the dispersion equation (\ref{trascendental}) the routine
\emph{NSolve} of Wolfram Mathematica 7 was used. We considered $a=1$. In the
case $A=1$ the routine \emph{NSolve }delivers one solution of
(\ref{trascendental}) $\lambda_{0}^{NSolve}=0.31902252414261895$. Application
of the SPPS method with $m=2000$ and $N=120$ gives us the value $\lambda
_{0}=0.31902252414254$. The agreement is up to the 12th digit. Taking $m=4000$
and $N=180$ we obtain still a better approximation $\lambda_{0}%
=0.319022524142619$. The agreement is up to the 14th digit.

In the case $A=4$ there are three eigenvalues. \emph{NSolve} delivers the
following values $\lambda_{0}^{NSolve}=0.41262411401896715$, $\lambda
_{1}^{NSolve}=2.8945478628320327$ and $\lambda_{2}^{NSolve}=3.749624961605374$%
. Application of the SPPS method with $m=2000$ and $N=100$ gives us the values
$\lambda_{0}=0.412624114002$, $\lambda_{1}=2.8945478628329$ and $\lambda
_{2}=3.7496249616095$, and with $m=4000$ and $N=180$: $\lambda_{0}%
=0.4126241140179$, $\lambda_{1}=2.89454786283226$ and $\lambda_{2}%
=3.7496249616045$.

\section{Conclusions\label{Sect Conclusions}}

We presented a review of recent research and applications of spectral
parameter powers series (SPPS) representations for solving initial and
boundary value problems as well as spectral and related problems for
Sturm-Liouville equations. Application of the SPPS approach allows one to
obtain explicit analytic forms of characteristic equations for a variety of
problems. Approximation of these equations represents a powerful, universal
and accurate numerical method highly competitive with the best purely
computational techniques. The SPPS method is algorithmically simple and can be
easily implemented using available routines of such environments for
scientific computing as Matlab.

\medskip

\noindent\textbf{Acknowledgments}

\noindent This research was partially supported by CONACYT, Mexico via the
research project 50424.

\end{document}